\documentclass[preprint,preprintnumbers,amsmath,amssymb]{revtex4}
\usepackage{graphicx}
\usepackage{epstopdf}
\usepackage{dcolumn}
\usepackage{bm}
\usepackage{color}
 \usepackage{array,multirow,graphicx}

\begin{document}

\title{Electronic, magnetic and optical properties of MnPX$_\mathbf{3}$ (X = S, Se) monolayers with and without chalcogen defects: A first-principle study}

\author{Juntao Yang,$^{1,2}$ Yong Zhou,$^{1}$ Qilin Guo,$^{1}$ Yuriy Dedkov,$^{1,}$\footnote{E-mail: dedkov@shu.edu.cn} and Elena Voloshina$^{1,}$\footnote{E-mail: voloshina@shu.edu.cn}}

\affiliation{$^1$Department of Physics, Shanghai University, 99 Shangda Road, 200444 Shanghai, China}
\affiliation{$^2$School of Science, Hubei University of Automotive Technology, 167 Checheng West Road, Shiyan City, 442002 Hubei, China}

\date{\today}

\begin{abstract}
Based on density functional theory (DFT), we performed first-principle studies on the electronic structure, magnetic state and optical properties of two-dimensional (2D) transition-metal phosphorous trichalcogenide MnPX$_3$ (X=S and Se). The calculated interlayer cleavage energies of MnPX$_3$ monolayers indicate the energetic possibility to be exfoliated from bulk phase, with good dynamical stability confirmed by the absence of the imaginary contributions in the phonon spectra. MnPX$_3$ monolayers are both N\'eel antiferromagnetic (AFM) semiconductors with direct band gaps falling into the visible optical spectrum. Magnetic interaction parameters were extracted within the Heisenberg model to investigate the origin of the AFM state. Three in-plane magnetic exchange parameters play important role in the robust AFM configuration of Mn ions. The N\'eel temperatures ($T_\mathrm{N}$) were estimated by means of Monte Carlo simulations, obtaining the theoretical $T_\mathrm{N}$ of $103$\,K and $80$\,K for 2D MnPS$_3$ and MnPSe$_3$, respectively. With high spin state, Mn ions arranged in honeycomb lattices, the spin-degenerated band structures exhibit valley polarisation and was investigated in different biaxial in-plain strains, considering the spin-orbital coupling (SOC). 2D MnPX$_3$ monolayers show excellent performance on the optical properties, and the absorption spectra were discussed in detail to find the transition mechanism. Different amount and configuration of chalcogen vacancy were introduced into the MnPX$_3$ monolayers, and it is found that the electronic structures are heavily affected depending on the vacancy geometric structure, leading to different magnetic state and absorption spectra of defected MnPX$_3$ systems.
\end{abstract}

\maketitle

\section{Introduction}

Two-dimensional (2D) van der Waals materials have been recently extensively explored,\cite{Geim:2009,Butler:2013ha,Geim:2014hf,Bhimanapati:2015bo,Lin:2018gi} owing to their unique electronic structures and interesting physical and chemical properties, since the first experiments on the demonstration of the extraordinary transport properties of graphene.\cite{Geim:2007a,Novoselov:2011} For several decades, the family of 2D crystals have grown considerably with new additions such as boron nitride (BN),\cite{Zhang:2017jx} black phosphorus (BP)\cite{Li:2014gf} and transition metal dichalcogenides (TMDs).\cite{Han:2018kb} Recently, a series of 2D transition metal trichalcogenides (TMTs) MPX$_3$ (M = V, Cr, Mn, Fe, Co, Ni and Zn; X = Se and Se) has gained many investigations over their synthesis, optical and electrical properties connected with weak interlayer van der Waals interactions.\cite{Kuo:2016ksa,Lee:2016jf,Susner:2017in,Wang:2018dha,Jenjeti:2018fu,Kim:2018gj,Gusmao:2019bt} Bulk 3D MPX$_3$ compounds can be prepared via chemical vapour deposition (CVD)\cite{Wang:2018dha} and chemical vapour transport (CVT)\cite{Du:2016ft} methods with high crystal quality. By mechanical exfoliation and chemical intercalation, the 2D monolayers of MPX$_3$ can be obtained with intermediate band gaps ranging from $1.3$\,eV to $3.5$\,eV,\cite{Du:2016ft} making them the ideal candidates for exfoliated 2D magnets and indicating their enhanced light absorption efficiency.\cite{Mukherjee:2016ku,Du:2016ft} DFT calculations predict that 2D MPX$_3$ monolayers exhibit a large variety of magnetic behaviours, including ferromagnetic (FM) metal, antiferromagnetic (AFM) semiconductors, and nonmagnetic (NM) insulators or metals, which can be effectively modulated via doping or lattice strain effects.\cite{Chittari:2016cda} These various magnetic functionalities of 2D MPX$_3$ can be employed for low dimensional spintronic and magnetroelectronic applications. The wide range of band gaps indicate that 2D MPX$_3$ compounds can also be considered as a promising candidates for optoelectronic, clean energy generation and related water splitting applications.\cite{Chu:2017hn,Du:2018dh}

Among the family of 2D MPX$_3$, AFM phase MnPX$_{3}$ compounds are the most interesting and widely focused materials due to prediction of some exciting properties, like visible-light catalytic activity,\cite{Zhang:2016kra} possibility to use in valleytronics,\cite{Li:2013ei} and doping-induced half-metallicity,\cite{Li:2014de} etc. Their magnetic structure and spin ordering, controlling through doping and strain effects have been theoretically investigated.\cite{Li:2014de,Sivadas:2015gq,Pei:2018jd} Notably, considered with the treatment of spin-orbital coupling, 2D MnPX$_{3}$ monolayers have been predicted with a spontaneous valley polarisation with degenerate spins.\cite{Li:2013ei} The magnetic behaviour and valley polarisation of 2D MnPX$_{3}$ monolayers can be controlled by transition metal substitutions and electronic coupling via heterostructures, resulting in FM, half-metallic and bipolar magnetic semiconductor.\cite{Zhong:2017gz,Pei:2017gc,Pei:2019gg} These strategies offers a practical avenue for exploring novel valleytronic devices which can be fabricated from 2D MnPX$_{3}$ monolayers.  

It is well know that defects (dopants, vacancies, interstitial atoms, etc.) play a significant role in tailoring of the 2D materials and the controllable modifications can lead to the drastic changes of the electronic, magnetic, optical, and transport properties. Although many density functional theory (DFT) calculations have been carried out on the electronic and magnetic properties, the optical properties of 2D MnPX$_{3}$ and the influence of chalcogen vacancies in these materials has been rarely investigated. Herein, we present an explicit investigation of the electronic structure of MnPX$_{3}$ monolayers in order to look insight the magnetic and optical properties in details. The influence of chalcogen vacancy on the electronic structure, magnetic and optical properties are also systemically studied and discussed in this work. 

\section{Computational details}

Spin-polarised DFT calculations based on plane-wave basis sets of $500$\,eV cutoff energy were performed with the Vienna \textit{ab initio} simulation package (VASP).\cite{Kresse:1994,Kresse:1996a,Kresse:1999} The Perdew-Burke-Ernzerhof (PBE) exchange-correlation functional\cite{Perdew:1996} was employed. The electron-ion interaction was described within the projector augmented wave (PAW) method\cite{Blochl:1994} with with Mn ($3p$, $3d$, $4s$), P ($3s$, $3p$), S ($3s$, $3p$) and Se ($4s$, $4p$) states treated as valence states. The Brillouin-zone integration was performed on $\Gamma$-centred symmetry reduced Monkhorst-Pack meshes using a Gaussian smearing with $\sigma = 0.05$\,eV, except for the calculation of total energies and densities of states (DOSs). For those calculations, the tetrahedron method with Bl\"ochl corrections\cite{Blochl:1994vg} was employed. The $12\times12\times4$ and $24\times24\times1$ $k$-meshes were used for the studies of bulk and monolayer MnPX$_3$, respectively, and the $12\times12\times1$ $k$-mesh was used for the $2\times2\times1$ supercells consisting of $4$-fold unit monolayers in case of vacancy studies.  In case of 2D MnPX$_3$, to ensure decoupling between periodically repeated layers, a vacuum space of $20$\,\AA\ was used. The convergence criteria for energy and force are set to $10^{-6}$\,eV and $0.005$\,eV\,\AA$^{-1}$, respectively.

The PBE+$U$ scheme\cite{Dudarev:1998vn} was adopted to properly describe the strongly correlated system of Mn\,$3d$ orbitals with the effective on-site Coulomb interaction parameter $U=5$\,eV.\cite{Li:2013ei,Franchini:2007gv} The HSE06 hybrid functional\cite{Heyd:2003eg} was also used for some systems with a $12\times12\times1$ $k$-mesh in order to get more accurate band gaps. Dispersion interactions were considered adding a $1/r^6$ atom-atom term as parameterised by Grimme (``D2'' parameterisation).\cite{Grimme:2006} 
 
The single layer lattice dynamical stability was determined using the first principles phonon calculations code PHONOPY\cite{Togo:2015jm} applying the finite displacement method\cite{Baroni:2001tn} within PBE+$U$. These phonons calculations were performed in $4\times4\times1$ supercells and very tight convergence criteria of as 1$0^{-8}$\,eV for energy and $0.1$\,meV\,\AA$^{-1}$ for force were used with $6\times6\times1$ $k$-mesh. Monte-Carlo simulations were performed within Metropolis algorithm\cite{METROPOLIS:1953vj} to estimate the $T_\mathrm{ N}$ value, using periodic boundary conditions with a series of superlattices containing different amount of magnetic sites. The optical spectra were calculated from the frequency dependent dielectric matrix after the electronic ground state has been determined.\cite{Gajdos:2006eh} The PYPROCAR code was used to plot the spin-textures.~\cite{Herath:2019qq} Crystal structure and charge density are visualised by VESTA.\cite{Momma:2011dd}

\section{Results and discussions}

\subsection{Electronic and magnetic structure of 3D MnPX$_3$}

\begin{figure}[h]
\centering
\includegraphics[width=0.46\textwidth]{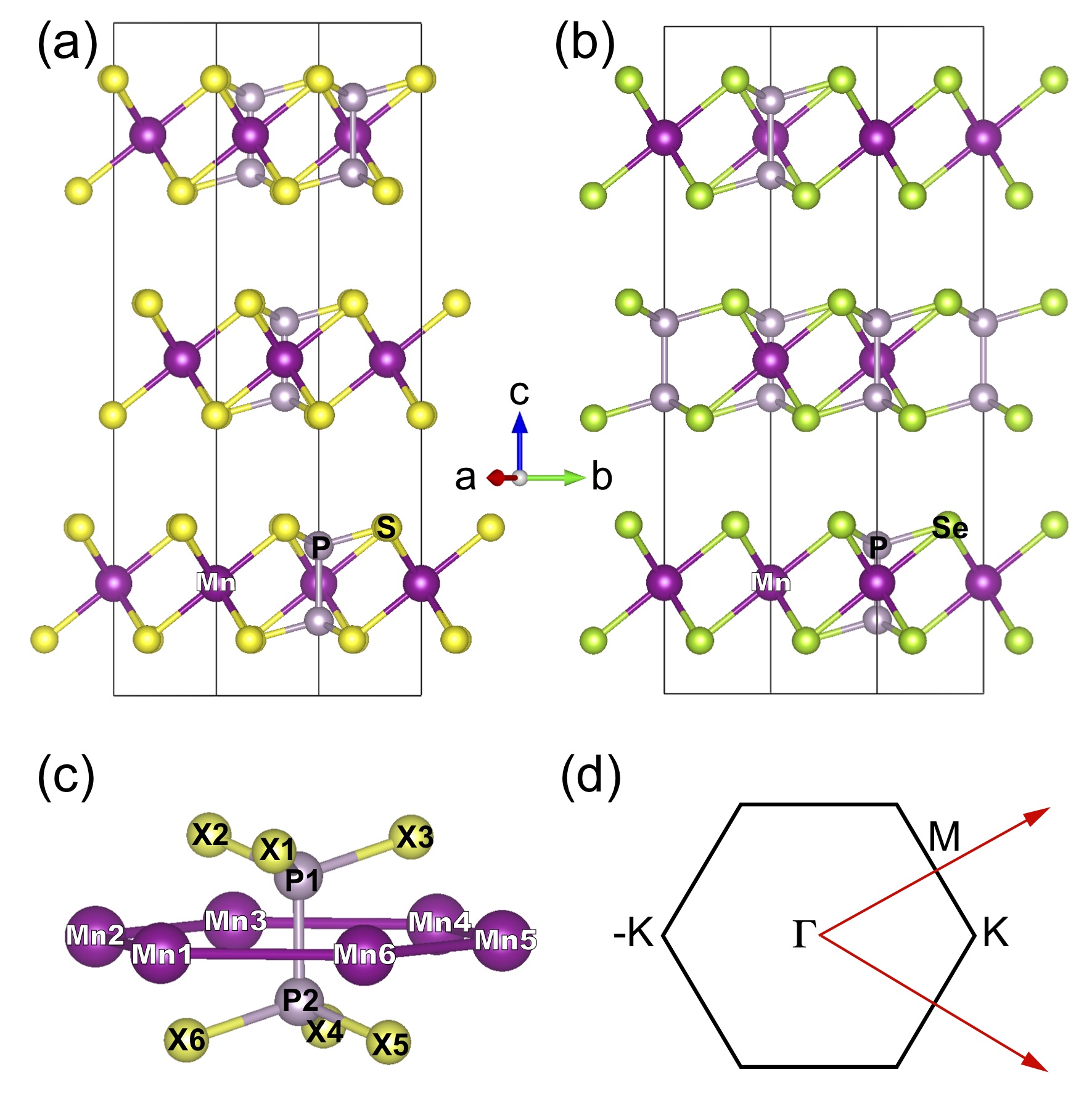} 	
\caption{Crystal structures of (a) 3D MnPS$_3$ and (b) 3D MnPSe$_3$. (c) Structure of the Mn-ions honeycomb lattice and centering [P$_2$X$_6$] bipyramid. (d) Sketch of the 2D Brillouin zone for the 2D hexagonal lattice with high symmetry $k$-points labeled.}
\label{CrystalStructure}
\end{figure}

The three-dimensional (3D) bulk MnPS$_3$ crystallises in $C2/m$ space group,\cite{Ouvrard:1985hi} while MnPSe$_3$ in $R\overline{3}$.\cite{WIEDENMANN:1981bs} Both of them can be represented in hexagonal unit cells as shown in Fig.~\ref{CrystalStructure} (a) and (b), respectively (for details, see ESI$^\dag$, Structural data for bulk MnPS$_3$ and Structural data for bulk MnPSe$_3$). Every unit cell contains three MnPX$_3$ single layers which has a $D_{3d}$ symmetry despite of the different 3D bulk space groups. The 2D MPX$_{3}$ layer is composed by two Mn$^{2+}$ ions which form hexagonal honeycomb lattice and one $[\textrm{P}_2\textrm{X}_6]^{4-}$ bipyramid built from a P-P dimer connected with two sulphur/selenium trimers. The P--P dimmer locates vertically across the centre of each honeycomb plane, with an in-plane twist of $60^{\circ}$ between the top and bottom trimmers as shown in Fig.~\ref{CrystalStructure} (c). The lattice parameters $a=b$ and $c$ of 3D MnPX$_3$ were fully relaxed in NM, FM and AFM magnetic states and they are listed in Table~\ref{3DLatticeParameters} with the respective total and relative energies. The energy difference
 $\Delta E= E^\mathrm{ AFM}-E^\mathrm{ FM}$ is $-164$\,meV  for 3D MnPS$_3$ and $-110$\,meV for 3D MnPSe$_3$, demonstrating that MnPX$_3$ bulk prefers AFM state than that of FM state. Especially, the NM sates of 3D MnPX$_3$ crystals show much higher energies than that of AFM one ($\Delta E> 25$\,eV), revealing that the NM state is strongly unfavourable in energy. The AFM ground state lattice parameters are in a good agreement with the experimental values.\cite{WIEDENMANN:1981bs,Ouvrard:1985hi} Clearly, each lattice parameter and monolayer thickness $d$ of MnPSe$_3$ are larger than the corresponding values of MnPS$_3$, caused by a larger ion radius of selenium than that of sulphur.

\begin{table}
\small
\caption{\ Relative to the lowest energy values ($\Delta E$, in eV) as well as lattice parameters (in \AA) for 3D MnPX$_3$ (X = S, Se) in different magnetic states. obtained with PBE+$\,U$+D2.}
\label{3DLatticeParameters}
  \begin{tabular*}{0.48\textwidth}{@{\extracolsep{\fill}}llllllll}
    \hline
X& State & $\Delta E$ & $a$  & $c$  &  $d$  &Mn-X  & Mn-Mn  \\
    \hline
S 	& NM		& $25.582$	& $5.789$		& $18.991$	& $3.065$		& $2.44$ 	& $3.34$ \\
			& FM		& $0.165$		& $6.070$		& $19.899$ 	& $3.314$		& $2.63$ 	& $3.50$\\
			& AFM 		& $0$		& $6.064$		& $19.893$ 	& $3.313$		& $2.62$ 	& $3.51$\\
& Expt$^a$	&			& $6.077$		& $20.388$	& 	\\
\hline
Se	& NM  		& $26.201$	& $6.144$ 	& $19.286$ 	& $3.379$ 	& $2.56$ 	& $3.55$ \\
			& FM  		& $0.113$		& $6.303$  	& $20.086$ 	& $3.491$ 	& $2.76$ 	& $3.69$ \\
			& AFM 		& $0$		& $6.398$  	& $20.082$ 	& $3.487$ 	& $2.76$ 	& $3.69$\\
& Expt$^b$ &		& $6.387$  	& $19.996$ 	&\\
    \hline
\end{tabular*}
$^a$Ref.~\citenum{Ouvrard:1985hi}\\
$^b$Ref.~\citenum{WIEDENMANN:1981bs}
\end{table}
\begin{table}
\small
\caption{\ Band gaps ($E_\mathrm{g}$, in eV) and Mn magnetic moments ($M$, in $\mu_\mathrm{B}$) obtained with different methods for 2D and 3D MnPX$_3$ systems. The available experimental values for band gaps are placed in parenthesis.}
\label{tablebandgap}
  \begin{tabular*}{0.48\textwidth}{@{\extracolsep{\fill}}llllllll}
\hline
		&\multicolumn{2}{c}{PBE }    &\multicolumn{2}{c}{PBE+$\,U$ }  &\multicolumn{2}{c}{HSE06} \\
		System  & $E_\mathrm{g}$ & $M$   & E$_\mathrm{g}$  & $M$  & E$_\mathrm{g}$  & $M$   \\ 
\hline	
3D MnPS$_3$ 		&$1.31$$^a$ 	& $4.24$	& $2.37$	&$4.60$	&$3.08$ ($3.0$$^b$)	&$4.48$ \\
2D MnPS$_3$ 		&$1.49$           		& $4.25$	& $2.50$	&$4.59$	&$3.25$  						&$4.49$  \\  
\hline
3D MnPSe$_3$	&$1.08$$^a$ 	& $4.25$	& $1.81$	&$4.58$	&$2.50$ ($2.5$$^b$) 	&$4.50$  \\
2D MnPSe$_3$ 	&$1.17$           		& $4.26$	& $1.84$	&$4.59$	&$2.56$  						&$4.50$   \\

\hline
\end{tabular*}
$^a$ Indirect\\
$^b$ Ref.~\citenum{Du:2016ft}
\end{table}
\begin{figure}[h]
\centering
\includegraphics[width=0.46\textwidth]{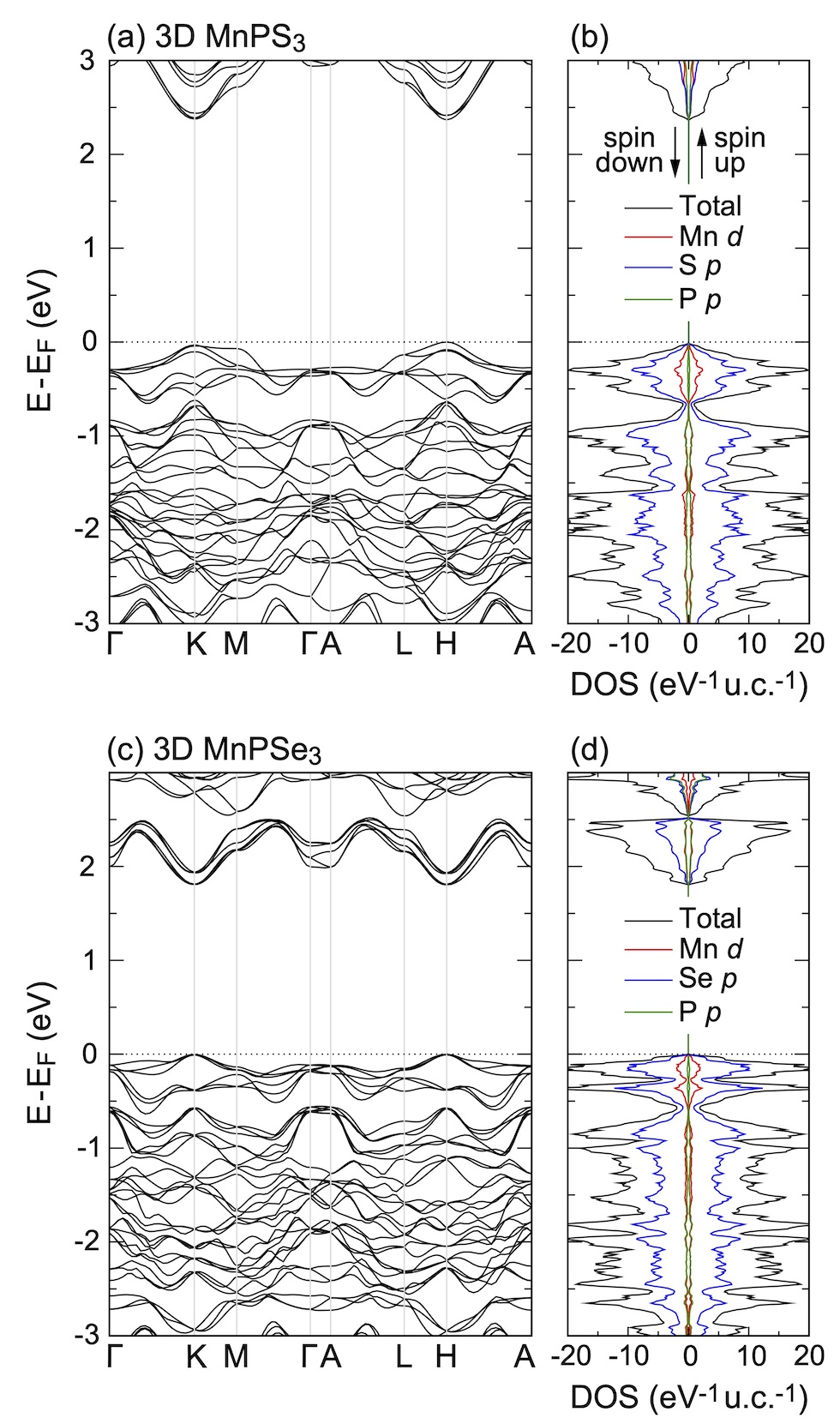} 	
\caption{Band structures, total and partial DOS of (a,b) 3D MnPS$_3$ and (c,d) 3D MnPSe$_3$, obtained with PBE+$\,U$.}
\label{3D-BandsDOSs}
\end{figure}

An interesting magnetic property for the bulk state is the interlayer magnetic coupling between monolayers in the unit cell. The energy differences between different inter-layer magnetic coupling $\Delta E_\mathrm{ int}= E_\mathrm{ int}^\mathrm{ AFM}-E_\mathrm{ int}^\mathrm{ FM}$ are $0.5$\,meV and $-0.27$\,meV for bulk MnPS$_3$ and MnPSe$_3$, respectively. Consequently, the FM inter-layer coupling is preferred for bulk MnPS$_3$, but AFM is preferred for bulk MnPSe$_3$. The different signs of $\Delta E_\mathrm{ int}$ can be attributed to the different monolayer arrangement for the two bulks in different space group. The small $\Delta E_\mathrm{int}$ is due to large distance of about $7$\,\AA\, between adjacent Mn honeycomb layers, and the neighbouring chalcogen atoms cannot act as a media bridge to mediate the long-range superexchage interactions across the large van der Waals gap, thus, the total energy is not sensitive to the inter-layer magnetic coupling, as that of FePS$_3$.\cite{Wang:2016ez} 

Now the electronic structure of 3D MnPX$_3$ is briefly discussed. The band structure and DOS obtained by PBE+$U$ method are shown in Fig.~\ref{3D-BandsDOSs} (a-b) and (c-d) for 3D MnPS$_3$ and MnPSe$_3$, respectively. The band structures are spin-degenerated in AFM state, showing similar features for both bulk MnPX$_3$ materials. The upper valence bands (VBs) at $E-E_F>-1.0$\,eV are mainly formed by S/Se $p$ and partly by Mn $3d$ orbitals; the lower conduction bands (CBs), at $E-E_F< 3.5$\,eV for 3D MnPS$_3$ and $E-E_F<2.5$\,eV for MnPSe$_3$, are mostly composed of S/Se $p$ and the partial contributions from P $p$ orbitals are almost equal to that of Mn $3d$ states. The band gaps and the magnetic moments of Mn ions were extracted and listed in Table~\ref{tablebandgap}. The GGA-PBE results give indirect band gaps of $1.31$\,eV for 3D MnPS$_3$ and $1.08$\,eV for 3D MnPS$_3$, respectively. As well known, the GGA approximation usually underestimates the semiconductor band gap, thus, PBE+$U$ and HSE06 methods were further used to evaluate the band structure and result into direct band gaps at the K point of the Brillouin Zone (BZ). The band gaps obtained by PBE+$U$ method are $2.37$\,eV for bulk MnPS$_3$ and $1.81$\,eV for bulk MnPSe$_3$, respectively. Moreover, the HSE06 functional predicts more accurate band gaps for the bulk systems with values of $3.08$\,eV and $2.50$\,eV for MnPS$_3$ and MnPSe$_3$, which shows good agreement with the experimental values of $3.0$\,eV and $2.5$\,eV,\cite{Du:2016ft} respectively. The magnetic moments of Mn ions calculated by PBE+$U$ method are $4.60\,\mu_\mathrm{ B}$ for both 3D MnPS$_3$ and MnPSe$_3$ (Table~\ref{tablebandgap}) and they are larger than that obtained with other functional. These calculated values are in good agreement with the one obtained in the neutron diffraction experiments: $4.40\,\mu_\mathrm{ B}$\cite{Kurosawa:1983aa} for MnPS$_3$ and $4.74\,\mu_\mathrm{ B}$\cite{WIEDENMANN:1981bs} for MnPSe$_3$, revealing the high spin state of Mn$^{2+}$ ions.

\begin{figure}
\centering
\includegraphics[width=0.46\textwidth]{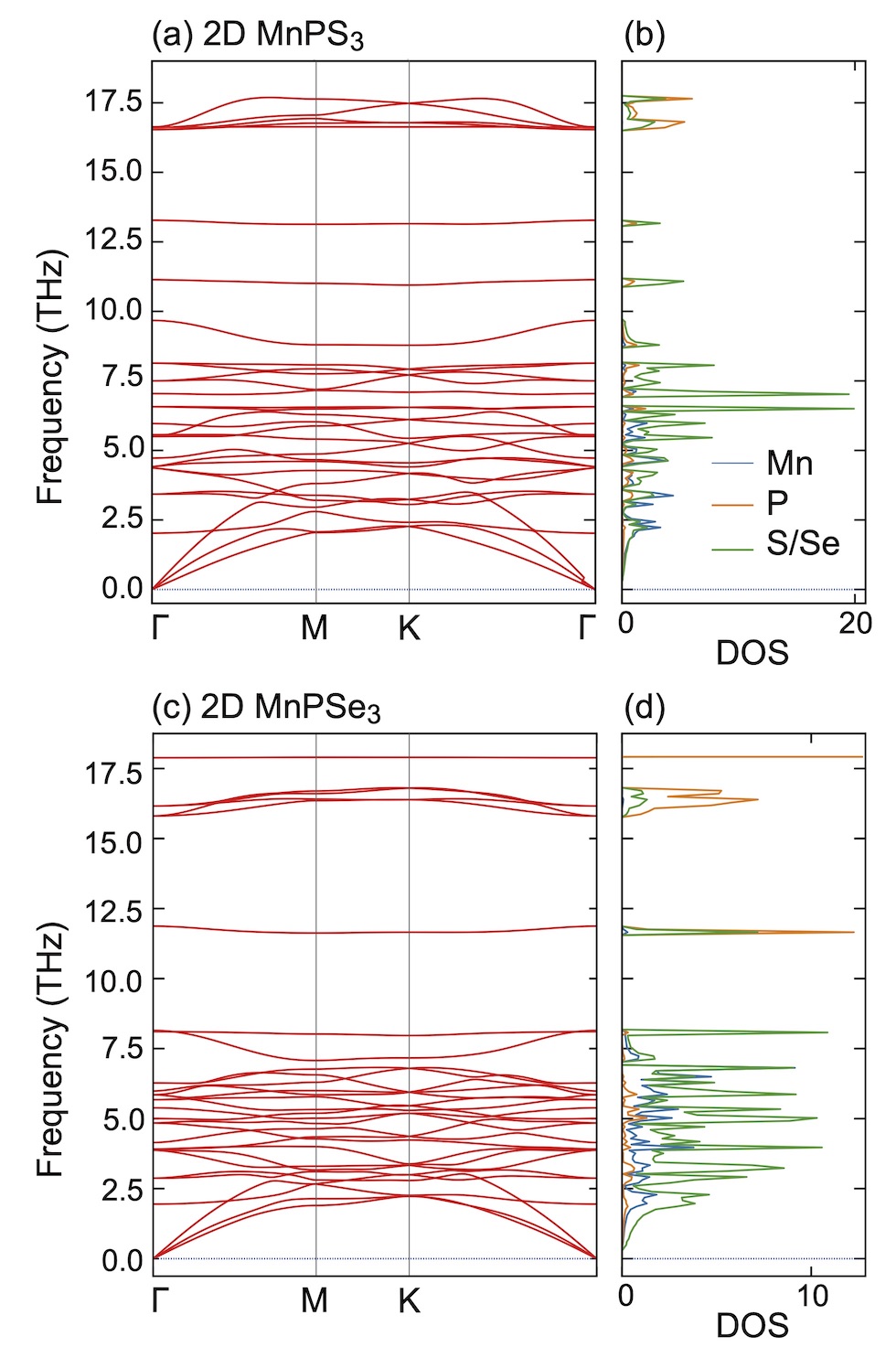} 	
\caption{Phonon dispersion spectra and the corresponding DOS for (a,b) 2D MnPS$_3$ and (c,d) 2D MnPSe$_3$.}
\label{Phonons}
\end{figure}

\subsection{Exfoliation energy and dynamical stability of 2D MnPX$_3$}

In order to evaluate the possibility of 2D MnPX$_3$ monolayer mechanical exfoliation from the bulk, the cleavage energy was calculated by $E_\mathrm{cl}=(E_{(d_0\rightarrow\infty)}-E_0)/{A}$, where $A$ is the in-plane area and $d_0$ is the van der Waals gap of bulk crystals. The theoretical value of $E_\mathrm{ cl}$ is $0.12\,\mathrm{J/m}^2$ for MnPS$_3$ and $0.23\,\mathrm{J/m}^2$ for MnPSe$_3$. The two $E_\mathrm{ cl}$ values are smaller than the experimentally estimated cleavage energy in graphite ($0.37\,\mathrm{J/m}^2$),~\cite{Zacharia:2004} indicating that the exfoliation of bulk MnPX$_3$ is feasible in experiments. The $E_\mathrm{ cl}=0.23\,\mathrm{J/m}^2$ value for MnPSe$_3$ is consistent with the previous theoretical results of $0.24\,\mathrm{J/m}^2$,\cite{Li:2014dea} whereas the $E_\mathrm{ cl}=0.12\,\mathrm{J/m}^2$ value for MnPS$_3$ is much lower than the previous calculated value about $0.26\,\mathrm{J/m}^2$.\cite{Du:2016ft} In other words, the $E_\mathrm{ cl}$ value for MnPS$_3$ is approximately twice smaller of the value calculated for MnPSe$_3$ as can be explained by two factors. The dominant one is that the Se ions offer larger van der Waals force than that of S ions, and it can be also confirmed by the comparison between $E_\mathrm{ cl}$ values for FePS$_3$ ($0.27\,\mathrm{J/m}^2$) and $R\overline{3}$ FePSe$_3$ ($0.38\,\mathrm{J/m}^2$).\cite{Du:2016ft} The second one is the different monolayer arrangement in the bulks between two bulk MnPX$_3$ compounds and can be proved by the $E_\mathrm{ cl}$ value of $0.16\,\mathrm{J/m}^2$ for fully relaxed MnPS$_3$ in the $R\overline{3}$ space group and $0.21\,\mathrm{J/m}^2$ for MnPSe$_3$ in the $C2/m$ space group.

To estimate the dynamical stability of MnPX$_3$ monolayers, the phonon spectra were calculated within the finite displacement method and the respective phonons dispersions are presented in Fig.~\ref{Phonons} along the high-symmetry directions of the hexagonal BZ. Importantly, no imaginary modes in the phonon dispersions are observed, confirming the dynamical stability of 2D MnPX$_3$, thus, they can be isolated in the experiments as freestanding layers. According to the phonons DOS given in right-hand side of the figure, the low frequency phonon bands below $8$\,THz for MnPX$_3$ are dominated by Mn--S/Se interactions, whereas the middle and high frequency branches above $8$\,THz originate from the internal molecular vibrations of the $[\textrm{P}_2\textrm{X}_6]$ group, consistent with the previous results.\cite{Hashemi:2017ds} According to the fact that the radius and mass of selenium atoms are larger compared to that of sulphur, and that the bond length of Mn--Se is longer than that of Mn--S, the phonon bands of MnPSe$_3$ are shifted to lower frequencies compared to that of MnPS$_3$.

\begin{table*}
\small
\caption{\ Total energy ($E_\mathrm{tot}$, in eV) and relative to the lowest energy ($\Delta E$, in meV) values for 2D MnPX$_3$ in the different magnetic states   obtained for $2\times 1\times 1$ supercell with PBE+$\,U$. Calculated and available experimental values for $T_\mathrm{N}$ (in K) are given in the last column.}
\label{2DMagneticEnergy}
  \begin{tabular*}{\textwidth}{@{\extracolsep{\fill}}lllllllllll}
\hline
	Monolayer 	& Energy   		 & FM     & AFM-N\'eel  & AFM-Zigzag  &  AFM-Stripy & $T_\mathrm{N}$\\
\hline
	MnPS$_3$	&$E_\mathrm{tot}$	& $-112.316$	& $-112.423$	& $-112.375$	& $-112.377$	&$103$ (Calc.)\\
				&$\Delta E$ 		& $107.14$	& $0$		&  $48.07$	& $46.51$		&$115$ (Expt.$^a$) \\
\hline
	MnPSe$_3$	&$E_\mathrm{tot}$	&$-104.620$ 	& $-104.697$	& $-104.666$   & $-104.659$ 	&$80$  (Calc.)  \\
				& $\Delta E$		 & $77.18$	& $0$		& $31.23$		& $37.53$   	&$74$ (Expt.$^b$)  \\

\hline
\end{tabular*}
$^a$Ref.~\citenum{MayorgaMartinez:2017cc}\\
$^b$Ref.~\citenum{WIEDENMANN:1981bs}
\end{table*}

\begin{figure*}
\centering
\includegraphics[width=0.92\textwidth]{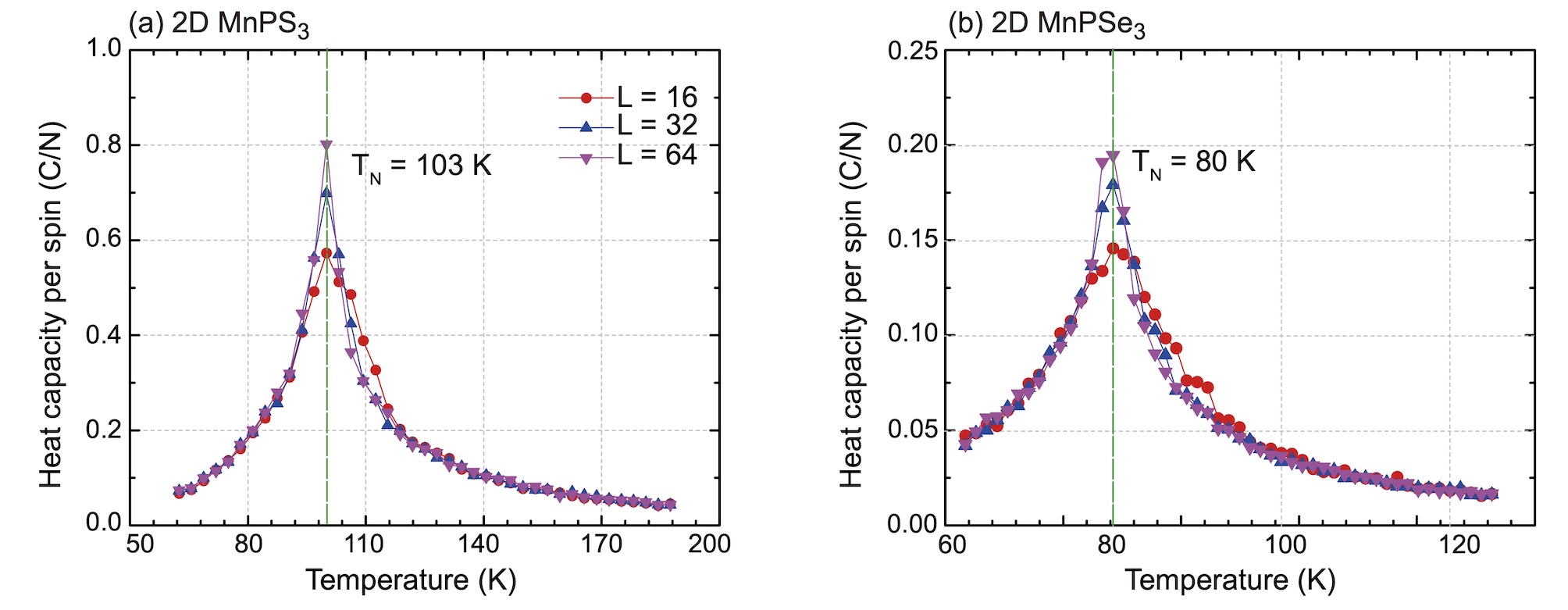}
\caption{Specific heat capacity with respect to temperature for (a) 2D MnPS$_3$ and (b) 2D MnPSe$_3$ for different lattice sizes used in the Monte Carlo simulations.}
\label{NeelTemperature}
\end{figure*}

\subsection{Magnetic properties of 2D MnPX$_3$}

Four possible magnetic configurations were investigated to evaluate the ground state of 2D MnPX$_3$ monolayers. The different spin configurations are FM, AFM-N\'eel, AFM-zig-zag and AFM-stripy (ESI$^\dag$, Fig.\,S1). To extract the total energies of different magnetic structures, we used four ordered spin states defined by using a $2\times1\times1$ supercell. The total energies of different spin configurations and the relative energy differences with respect to the AFM-N\'eel configuration calculated within the PBE+$U$ method are listed in Table~\ref{2DMagneticEnergy}. It can be seen that the lowest total energy is the AFM-N\'eel state for 2D MnPX$_3$ monolayer. In addition, NM configurations were also calculated, showing NM MnPX$_3$ monolayers are in semi-metallic state. Similar to that of 3D bulk unit systems, the NM state of 2D MnPX$_3$ monolayers also show much larger energies compared to the AFM ground state by $8.78$\,eV for MnPS$_3$ and by $8.87$\,eV for MnPSe$_3$, demonstrating that the MnPX$_3$ monolayers persist the AFM magnetism similar to the 3D bulk system. Accordingly, the Mn ions magnetic moments of 2D MnPX$_3$ are comparable to values for 3D bulk state, listed in Table.~\ref{tablebandgap}. Therefore one can conclude that 2D MnPX$_3$ is a robust intrinsic AFM monolayer.
\begin{figure}[h]
\centering
\includegraphics[width=0.46\textwidth]{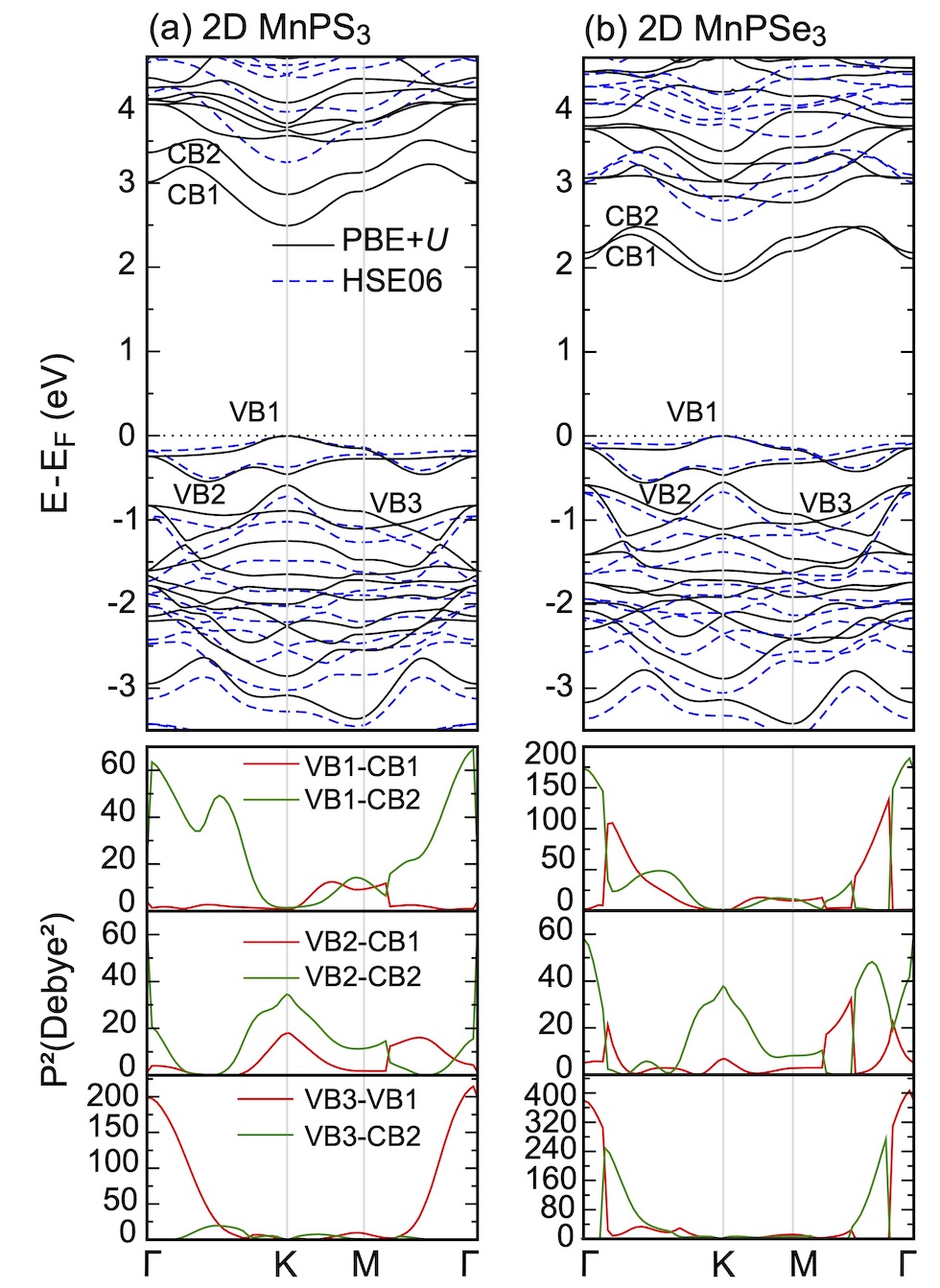}
\caption{PBE+$\,U$ calculated band structures of (a) 2D MnPS$_3$ and (b) 2D MnPSe$_3$, compared with that obtained with the HSE06 functional, respectively. The lower panels are the corresponding square of transition dipole moment matrix.}
\label{2DBandStructure}
\end{figure}

\vspace{-0.3cm}
To extract the exchange interaction parameters between Mn ions spins, the Heisenberg Hamiltonian was considered
\begin{equation}
H=\sum_{\langle i,j \rangle}J_1\vec{S}_i \cdot \vec{S}_j+\sum_{\langle \langle i,j \rangle \rangle}J_2\vec{S}_i \cdot \vec{S}_j+\sum_{\langle \langle \langle i,j \rangle \rangle \rangle}J_3\vec{S}_i \cdot \vec{S}_j,\nonumber
\end{equation}
where $\vec{S}_i$ is the total spin magnetic moment of the atomic site $i$, $J_{ij}$ are the exchange coupling parameters between two local spins. Considering one central Mn atom interacted with three nearest neighbours (NN, $J_1$), six next-nearest neighbours (2NN, $J_2$), and three third-nearest neighbours (3NN, $J_3$) Mn atoms. Here, the long-range magnetic exchange parameters ($J$) can be obtained by \cite{Sivadas:2015gq}
\begin{eqnarray}\nonumber
J_1=\frac{E_\mathrm{FM}-E_\mathrm{AFM\textrm{-}Neel}+E_\mathrm{AFM\textrm{-}zz}-E_\mathrm{AFM\textrm{-}str}}{8S^2},\\ \nonumber
J_2=\frac{E_\mathrm{FM}+E_\mathrm{AFM\textrm{-}Neel}-(E_\mathrm{AFM\textrm{-}zz}+E_\mathrm{AFM\textrm{-}str})}{16S^2},\\\nonumber
J_3=\frac{E_\mathrm{FM}-E_\mathrm{AFM\textrm{-}Neel}-3(E_\mathrm{ AFM\textrm{-}zz}-E_\mathrm{AFM\textrm{-}str})}{24S^2}.\nonumber
\end{eqnarray} 
where $S$ is the calculated magnetic moments of Mn ion, and $E_\mathrm{FM}$, $E_\mathrm{ AFM\textrm{-}Neel}$, $E_\mathrm{ AFM\textrm{-}zz}$ and $E_\mathrm{ AFM\textrm{-}str}$ are the total energies in FM, AFM-N\'eel, AFM-Zigzag and AFM-Stripy magnetic configurations, respectively.

Using the presented results one gets $J_1=0.65$\,meV,  $J_2=0.037$\,meV,  $J_3=0.20$\,meV for 2D MnPS$_3$ and $J_1=0.47$\,meV, $J_2=0.03$\,meV, $J_3=0.19$\,meV for 2D MnPSe$_3$, which are in excellent agreement with the results from the previous studies.\cite{Sivadas:2015gq} With all positive exchange parameters $J$, these results indicate that both MnPS$_3$ and MnPSe$_3$ monolayers are in robust AFM-N\'eel phase. The significant exchange interaction values indicate that the 2NN $J_2$ and 3NN $J_3$ exchange couplings make important contributions to the ground magnetic state, besides the NN  $J_1$. The NN exchange $J_1$ state comes from the competition between NN Mn-Mn direct exchange and Mn--X--Mn superexchange. The direct Mn-Mn interaction is always AFM according to the $d$ orbitals overlaps, while the Mn-X-Mn superexchange is always FM due to the Mn-X-Mn angle which is close to $90^{\circ}$ ($84.26^{\circ}$ for MnPS$_3$ and $84.03^{\circ}$ for MnPSe$_3$), as can be understandable from the well-known Goodenough-Kanamori-Anderson (GKA) rules.\cite{Goodenough:1955qqq,Kanamori:1960ki,Anderson:1959aaa} Because of the high spin Mn$^{2+}$ $d^5$ state and the short Mn-Mn distance ($3.51$\,\AA\ for MnPS$_3$ and $3.69$\,\AA\ for MnPSe$_3$), the AFM direct exchange wins the competition and dominate the value of $J_1$. The 2NN $J_2$ and 3NN $J_3$ exchange coupling parameters can be considered as super-superexchange interactions mediated by Mn--X$\cdots$X--Mn bridge, and the strong hybridisation between X n$p$ orbitals and Mn 3$d$ orbitals (see below) resulting in AFM interactions. According to the monolayer geometry, $J_3$ involves Mn--X1$\cdots$X3--Mn bridge with two X ions locate in the same chalcogen sub-layer and it is stronger than $J_2$ by Mn-X1$\cdots$X5-Mn with X ions locate in separate sub-layers. In general, the AFM-N\'eel ground state is mostly from a strong AFM direct exchanges between the Mn$^{2+}$ions for monolayer MnPX$_3$. 

On basis of Ising model, Monte Carlo simulations with periodic boundary conditions were performed to estimate the N\'eel temperatures of MnPX$_3$ monolayers. The three exchange parameters $J_1$, $J_2$ and $J_3$ were used in a series of superlattice $L\times L$ ($L=16, 32, 64$) containing a large amount of magnetic sites to accurately evaluate the value. Upon the heat capacity $C_v(T)=(\langle E^2\rangle - \langle E\rangle^2)/k_\mathrm{B}T^2$ reaches the equilibrium state at given temperature, the $T_\mathrm{ N}$ can be extracted from the peak of the specific heat profile. The specific heat capacity per spin as a function of temperature are plotted in Fig.~\ref{NeelTemperature}. From the simulated $C_v(T)$ curves, one can find the peak ascend steepening with the lattice size increases. The accurately estimated $T_\mathrm{ N}$ value is $103$\,K for 2D MnPS$_3$ and $80$\,K for 2D MnPSe$_3$, which is in excellent agreement with the experimental values of $100$\,K\cite{MayorgaMartinez:2017cc} evaluated from the susceptibility data for MnPS$_3$ and $74$\,K\cite{WIEDENMANN:1981bs} revealed by neutron diffraction experiments for MnPSe$_3$ in the bulk phase.

\subsection{Electronic structures of 2D MnPX$_3$}

Having studied the magnetic ground state, the electronic properties of 2D MPX$_3$ were investigated. The band structures calculated by PBE+$U$ and HSE06 method are presented in Fig.~\ref{2DBandStructure} (a) and (b) for 2D MnPS$_3$ and MnPSe$_3$, respectively. As expected, both monolayers demonstrate the semiconductor behaviour. The band gaps calculated by different functionals are given in Table.~\ref{tablebandgap}. With the value of $3.25$~\,eV for 2D MnPS$_3$ and $2.56$~\,eV for 2D MnPSe$_3$ obtained by HSE06 functional, the gap is little larger than the corresponding value of 3D bulk  due to the quantum confinement effect when going from 3D to 2D.  Unlike that of 3D systems, the pure PBE functional gives direct band gaps at K point as the other considered approaches (PBE+$U$ and HSE06). As a consequence, 2D MPX$_3$ can act as good candidates for 2D magnets semiconductors with localised magnetic moment on the Mn ions.

The upper VBs (above $-0.5$\,eV) are composed of two bands, which obtained by PBE+$U$, almost overlap that of HSE06 results. The other HSE06 VBs move towards lower energy range compared to that of PBE+$U$ bands. One can see that the HSE06 CBs all move up to higher energy range compared to that of PBE+$U$ bands, thus the HSE06 band gap are the best consistent with the experimental values. However, the bands dispersion of HSE06 is almost similar to that of PBE+$U$, that is to say, the main characters of the band structures are almost correctly described by PBE+$U$ method. Therefore, we focused on the total DOS and partial DOS of 2D MnPX$_3$ obtained within the PBE+$U$ method.

\begin{figure*}[h]
\centering
\includegraphics[width=0.92\textwidth]{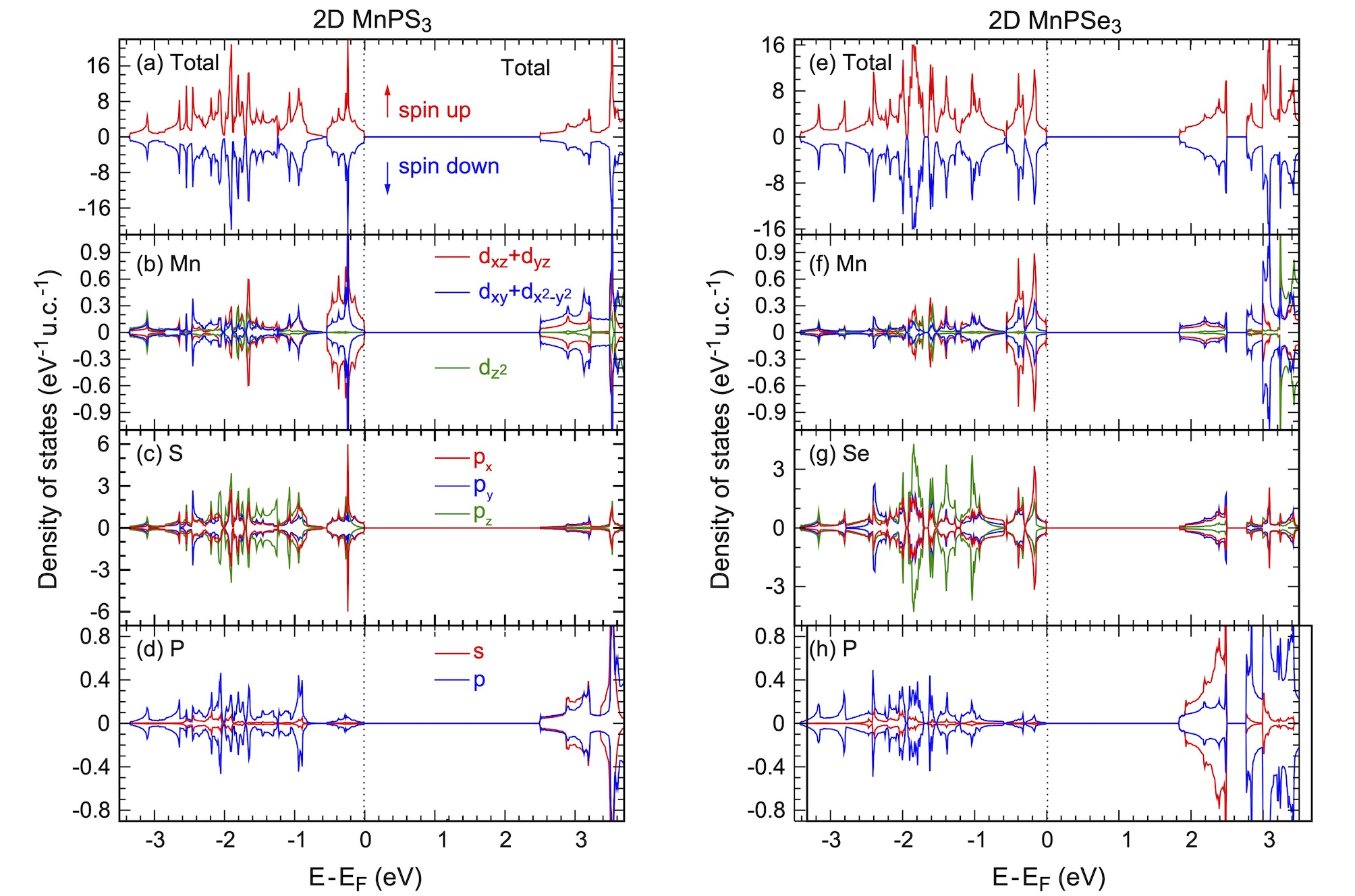}
\caption{Total and partial DOS states of (a) 2D MnPS$_3$ and (b) 2D MnPSe$_3$ obtained with PBE+$U$.}
\label{2D-DOS}
\end{figure*}
\vspace{-0.05cm}

The total DOS and partial DOS are shown in Fig.~\ref{2D-DOS}: (a) for MnPS$_3$ and (b) 2D MnPSe$_3$, respectively. Due to the AFM ground state, the DOS is identical for spin up and spin down channels. Since the trigonal anti-prismatic MnX$_6$ octahedron is under $D_{3d}$ symmetry in both monolayers, the Mn\,$3d$ orbitals can be decomposed into a single $a_1$ ($d_{z^2}$) orbital, and two $2$-fold degenerate $e_1$ ($d_{xz}$, $d_{yz}$) and $e_2$ ($d_{xy}$, $d_{x^2-y^2}$) orbitals. The five $d$ electrons  occupy only one spin channel, leading to the high spin state of Mn ions, due to the strong crystal field effect. The top of the valence band VB is mainly dominated by the hybridisation between Mn\,$d-e_1$, S/Se\,$p_x$ and $p_y$ orbitals, as shown in Figs.~\ref{2D-DOS} (b, c) and (f, g). Moreover, the strong hybridisation between Mn\,$d-e_1,e_2$ orbitals and S/Se\,$p_x,p_y$  orbitals can been clearly seen in the whole energy range, confirming the superexchange interactions between the Mn\,$d$ orbitals mediated by the S/Se\,$p$ orbitals. The bottom of the conduction band CB is derived from Mn\,$d-e_1,e_2$, S/Se\,$p_x$, $p_y$ and P\,$p$, $s$ orbitals.  Compared with that of 2D MnPS$_3$, the contributions from the P\,$s$ orbitals to the two lower CBs are stronger than that of P $p$ states for 2D MnPSe$_3$, as shown in Figs.~\ref{2D-DOS} (d) and (h), respectively. Clearly, a $0.25$\,eV sub-gap appears at $2.6$\,eV in CBs for MnPSe$_3$ above the two lower CBs. In addition, from $-0.5$\,eV to the Fermi level of the upper VBs, the hybridisation effect between Mn\,$3d$ and S\,$p$ orbitals of 2D MnPS$_3$, as reflected by one main sharp peak, is stronger than that of 2D MnPSe$_3$ where two main sharp peaks are observed, resulting in more localised electron properties for MnPS$_3$ than that for MnPSe$_3$.

\vspace{-0.02cm}

The different electronic structures of MnPX$_3$ monolayers can be attributed to two factors: (i) the stronger crystal field effect in MnS$_6$ octahedron than that of MnSe$_6$ one due to the larger atomic radius and bond lengths of Se ions than that of S ions, as listed in Table \ref{3DLatticeParameters}; (ii) S ions have stronger electronegativity than Se ions, thus, the onsite energy of Se $p$ and $s$ orbitals is closer to that of Mn $d$ states compared to the S ions. As a consequence, the $d-p$ hybridisation in MnPS$_3$ is stronger than that of MnPSe$_3$, resulting to the larger band gap and more itinerant electrons of MnPS$_3$ than that of MnPSe$_3$. Moreover, the two factors can lead to stronger Mn-X$\cdots$X-Mn mediated super-superexchange interaction of MnPS$_3$ than that of MnPSe$_3$, which can clarify that MnPS$_3$ has higher $T_\mathrm{ N}$ than MnPSe$_3$. 
\begin{figure}[h]
\centering
\includegraphics[width=0.46\textwidth]{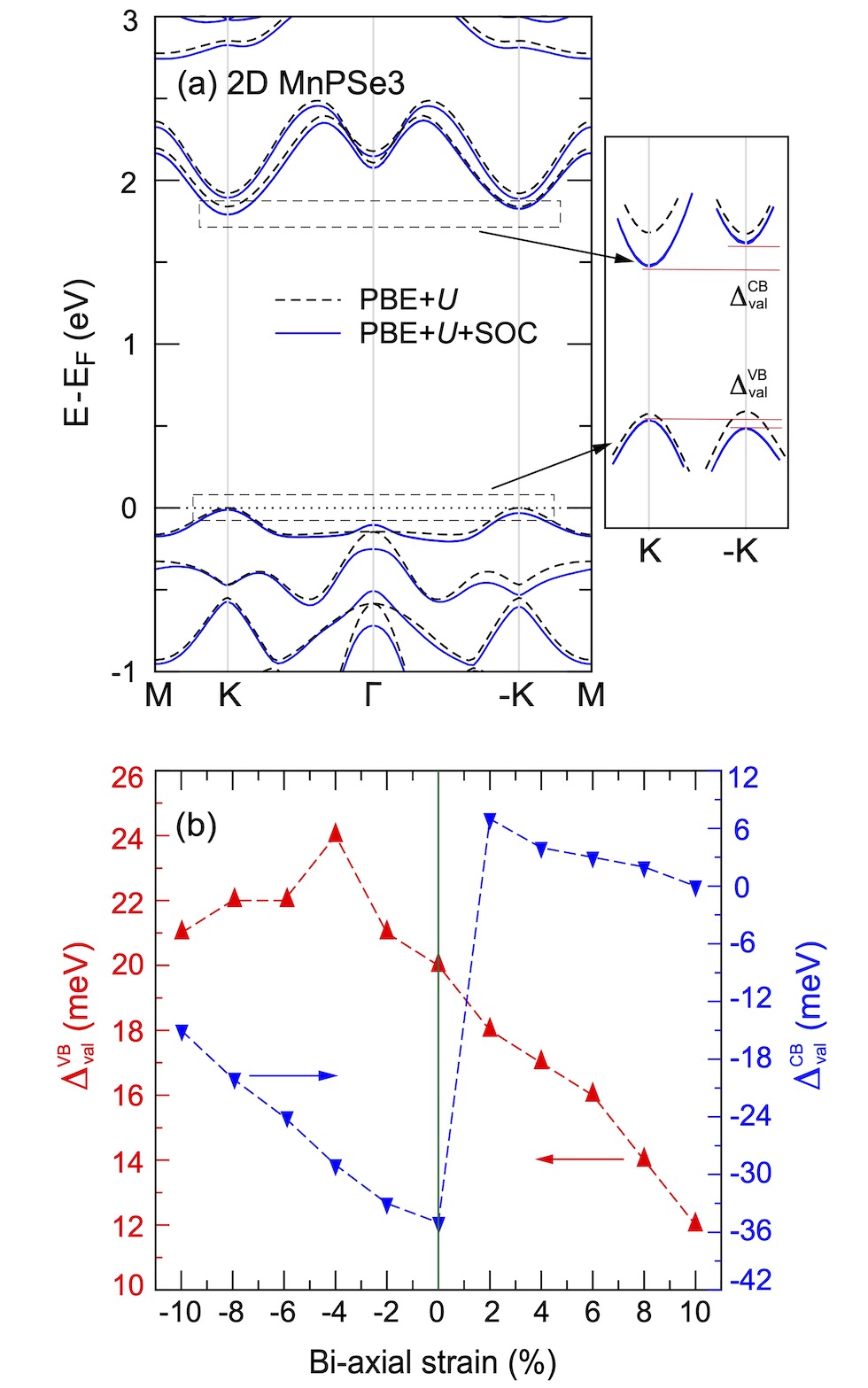}
\caption{(a) Band structure of 2D MnPSe$_3$ calculated with and without SOC within the PBE+$U$ method. The schematic diagram of bands at $\mathrm{K}$ and $\mathrm{-K}$ valleys is presented on the right-hand side of (a).  (b) The effect of in-plain bi-axial strain to valley polarisation $\Delta_\mathrm{val}^\mathrm{CB/VB}$ at the $\mathrm{K}$ and $\mathrm{-K}$ point.}
\label{2DBandStructureSOC}
\end{figure}

As a N\'eel AFM semiconductor is constructed with honeycomb lattice, a spontaneous valley polarisation along with degenerate spins can be realised in 2D MnPX$_3$ monolayers.\cite{Li:2013ei} In order to study the valley polarisation, 2D MnPX$_3$ band structures were further investigated taking into account the spin-orbital coupling (SOC) (Fig.\,S2 in Supplementary material). The valley polarisation of 2D MnPSe$_3$ is more significant than that of 2D MnPS$_3$ (discussed below), herein we only present the band structure of MnPSe$_3$ monolayer with SOC taken into account (Fig.~\ref{2DBandStructureSOC} (a)). The respective valley polarisation can be estimated by the energy differences between in the uppermost VB and lowest CB at the $\mathrm{K}$ and $\mathrm{-K}$ point of BZ, defined as $\Delta_\mathrm{ val}^\mathrm{CB/VB}=E_\mathrm{ K}^\mathrm{ CB/VB}-E_\mathrm{ -K}^\mathrm{ CB/VB}$. The spontaneous valley degeneracy splittings of about $\Delta_\mathrm{ val}^\mathrm{CB/VB}=7/-3$\,meV for MnPS$_3$ and $20/-35$\,meV for MnPSe$_3$ appear for the uppermost VB band and lowest CB bands, respectively, which are consistent with previous studies.\cite{Li:2013ei,Pei:2019gg} Due to the valley polarisation, the band gap slightly reduces to $2.49$\,eV for 2D MnPS$_3$ and $1.80$\,eV for 2D MnPSe$_3$, but remains direct at the $\mathrm{K}$ point.  Caused by the SOC effect, the upper VBs show large energy-splittings about $0.2$\,eV at the $\Gamma$ point, compared to that of PBE+$U$ bands, which demonstrate the twofold energy-degeneracy for the occupied VBs. Except for the $\Gamma$ point, the SOC band dispersion shows similar features as that of PBE+$U$ with small energy shifts along the high symmetry $\mathrm{k}$-path. Due to the time-reversal symmetry, the total magnetic moments are $0\,\mu_\mathrm{ B}$ for any direction spin components for both 2D MnPX$_3$ monolayers, as demonstrated by the SOC bands which do not show any spin-splitting in the momentum space.

Here, we also studied the effect of the biaxial in-plain strain in the MnPSe$_3$ monolayer on the valley polarisation. The results of $\Delta_\mathrm{ val}^\mathrm{ VB}$  and  $\Delta_\mathrm{ val}^\mathrm{ CB}$ with respect to the strain are plotted in Fig.~\ref{2DBandStructureSOC} (b) with positive and negative strains representing tensile and compressive effects, respectively. The tensile strain will enlarge the Mn-Mn bonds length while the compressive effect will enlarge the $d-d$ direct AFM interactions. It is found that the tensile strains reduce valley polarisation $\Delta_\mathrm{ val}^\mathrm{ VB}$, whereas the compressive strains slightly increase the value. For $\Delta_\mathrm{val}^\mathrm{CB}$, both tensile and compressive strains reduce the absolute value, and interestingly the tensile strain will change $\Delta_\mathrm{ val}^\mathrm{ CB}$ sign from negative to positive. As a result, the SOC band gap remains direct character under compressive strain but change to indirect feature under tensile effect (uppermost VB at $\mathrm{K}$ point band and lowest CB at $\mathrm{-K}$ point). The strain effect to the band gap of SOC can be evaluated as $\Delta E_\mathrm{ g}=E_\mathrm{ g}(\textrm{K})-E_\mathrm{ g}(-\textrm{K})=|\Delta_\mathrm{ val}^\mathrm{ CB}|+|\Delta_\mathrm{ val}^\mathrm{ VB}|$. The estimated maximum value of $\Delta E_\mathrm{ g}$ is $55$\,meV in equilibrium state. That is to say, the valley polarisation of pristine MnPSe$_3$ monolayers most easily to be detected in experiment without in-plane strain which offers an undesirable influence on the valley polarisation.

\subsection{Optical properties of 2D MnPX$_3$}

Besides the interesting magnetic properties, the 2D MnPX$_3$ monolayers also show significant optical performance with band gaps in the fundamental range. The linear optical properties can be obtained from the frequency-dependent complex dielectric function as $\varepsilon(\omega)=\varepsilon_1(\omega)+i\varepsilon_2(\omega)$, where $\varepsilon_1(\omega)$ and $\varepsilon_2(\omega)$ are the real and imaginary parts, respectively, of the dielectric function at the photon energy $\omega$.\cite{Gajdos:2006eh} To describes the decay of light intensity spreading in unit distance in medium, the absorption coefficient $\alpha(\omega)$ can be calculated from the dielectric function as~\cite{Zhang:2016kra,Wang:2014bf}
\begin{eqnarray}\nonumber
\alpha(\omega)=\sqrt{2}\omega\left\{\frac{\sqrt{\varepsilon_1^2(\omega)+\varepsilon_2^2(\omega)}-\varepsilon_1(\omega)}{2}\right\}^{1/2}.
\end{eqnarray}

Since the point group of N\'eel AFM state is $D_{3d}$, 2D MnPX$_3$ monolayers process inversion symmetry which will introduce parity-forbidden transitions between VBs and CBs.\cite{Meng:2017bq} In order to investigate the effect of symmetry-induced parity-forbidden transitions to the optical absorption properties, the transition dipole moment (TDM) defined by the electric dipole moment were calculated. For a single, non-relativistic particle of mass $m$, the TDM can be denoted in terms of the momentum operator $\mathbf{p}$,\cite{Wang:2019wp}
\begin{eqnarray}\nonumber
P_{a\rightarrow b}= \langle\psi_b|\mathbf{r}|\psi_a\rangle=\frac{i\hbar}{(E_b-E_a)m}\langle\psi_b|\mathbf{p}|\psi_a\rangle
\end{eqnarray}
TDM $P_{a\rightarrow b}$ means a transition where a single charged particle changes from initial state $|\psi_a \rangle$ in an occupied band to final state $|\psi_b\rangle$ in an empty band with energy $E_a$ and $E_b$ at its position $\mathbf{r}$. In general, the sum of square of TDM ($P^2$ in unit of $\textrm{Debye}^2$) implies the transition probabilities between the initial and final states. 

\begin{figure*}
\centering
\includegraphics[width=0.92\textwidth]{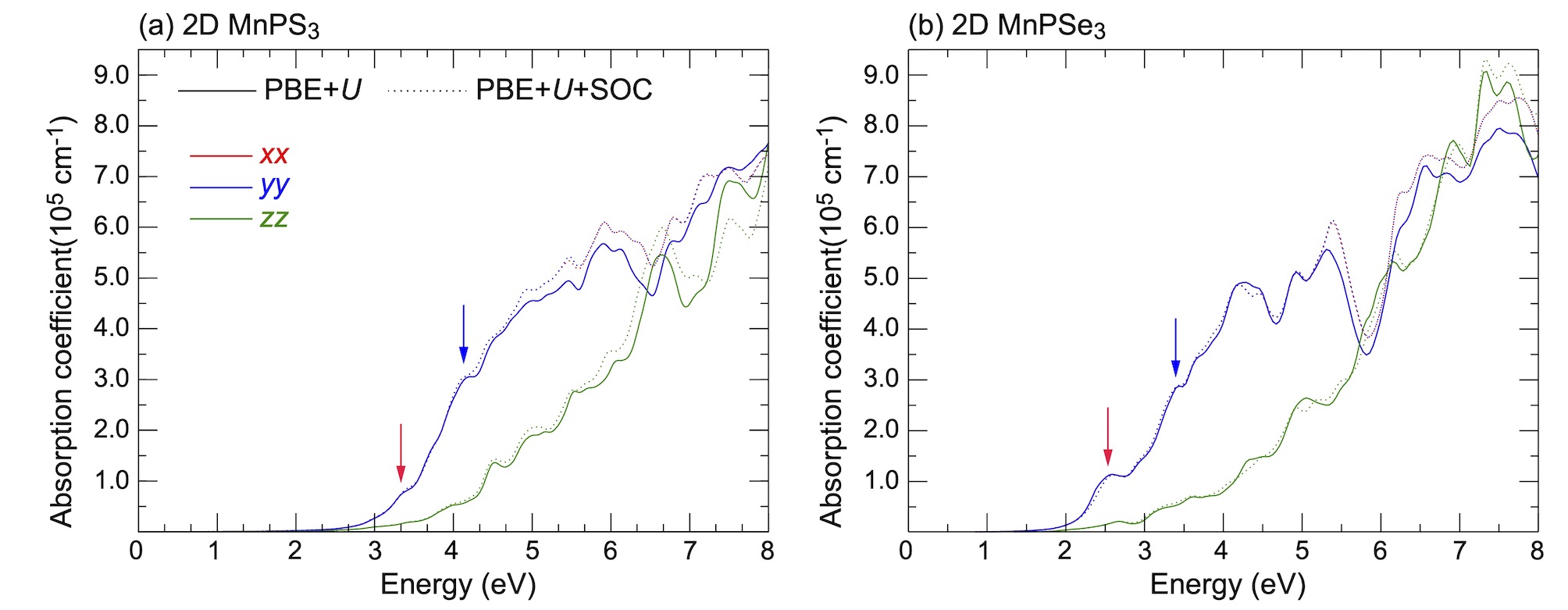}
\caption{The absorption spectra of (a) 2D MnPS$_3$ and (b) 2D MnPSe$_3$ considered with and without SOC effect within the PBE+$\,U$ method.}
\label{OpticsAbsorption}
\end{figure*}
\begin{figure*}
\centering
\includegraphics[width=0.92\textwidth]{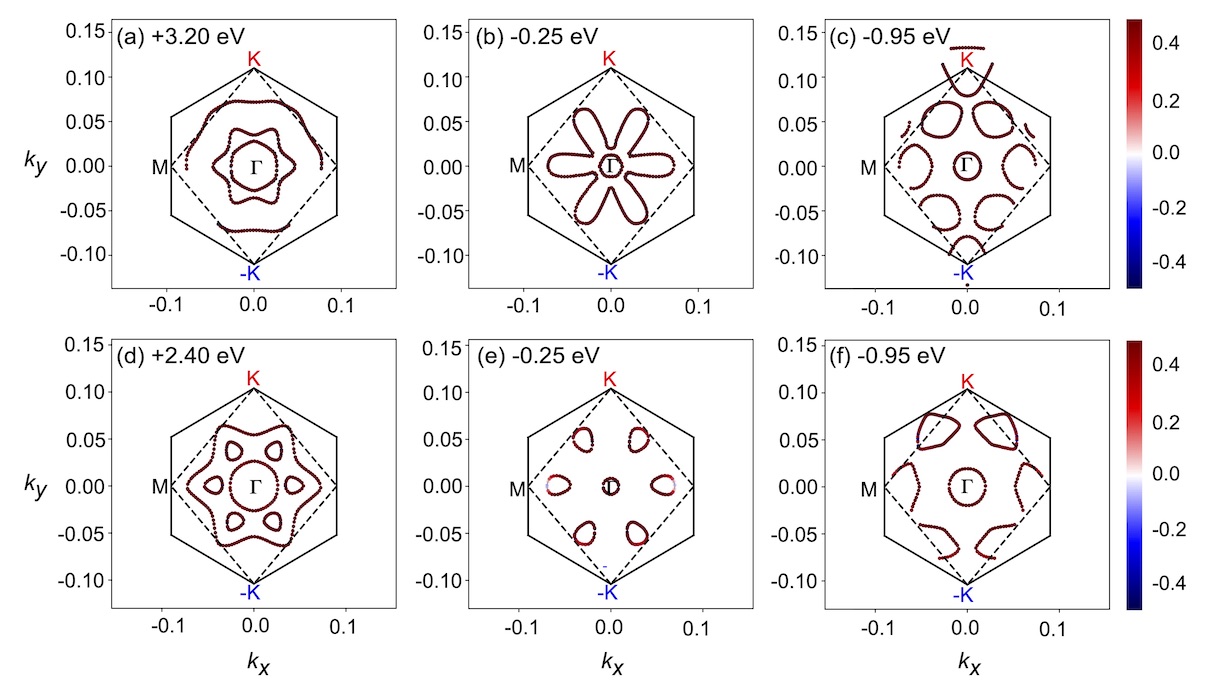}
\caption{Spin textures of 2D MnPS$_3$ (a-c) and 2D MnPSe$_3$ (d-f) shown in the $k_x-k_y$ plane cantered at the $\Gamma$ point at different energies $E-E_F$. The projection of the spin on the $z$ axis from negative to positive is colour coded with the blue-red scale. The solid line hexagonal is the BZ and the broken line is the edge of spin-textures.}
\label{SpinTexture2D}
\end{figure*}

\begin{table*}
\caption{\ Formation energy ($E_\mathrm{f}$, in eV), Mn ions magnetic moments (orbital and total, $M$, in $\mu_\mathrm{B}$) , and band gaps ($E_\mathrm{g}$, in eV) for spin up/down electrons obtained for different defected 2D MnPX$_3$ systems with PBE+$\,U$. }
\label{vacancytable}
  \begin{tabular*}{\textwidth}{@{\extracolsep{\fill}}lllllllllll}
\hline
Vacancy   &$E_\mathrm{f}$ & $M_\mathrm{Mn1}/M_\mathrm{Mn6}$ &  $M_\mathrm{Mn2}/M_\mathrm{Mn5}$  & $M_\mathrm{Mn3}/M_\mathrm{Mn4}$ & $M_{s}$/$M_{p}$/$M_{d}$ & $M_\mathrm{tot}$ &$E_\mathrm{g}^\mathrm{up}/E_\mathrm{g}^\mathrm{down}$\\
\hline
V$_\mathrm{S}$@1L  	&$1.41$  &$-4.584$/$4.583$ &$4.583$/$-4.583$ &$-4.587$/$4.587$ &$-0.001$/$0.003$/$-0.001$ & $0.001$ & $2.39$/$2.38$\\
V$_\mathrm{S2}$@1L  	&$1.65$  &$-4.584$/$4.584$ &$4.593$/$-4.592$ &$-4.587$/$4.586$ &$-0.001$/$0.003$/$0.000$  & $0.002$ & $1.62$/$1.63$ \\  
V$_\mathrm{S2}$@2L  	&$1.47$  &$-4.587$/$4.587$ &$4.579$/$-4.579$ &$-4.587$/$4.587$ &$0.000$/$0.000$/$0.000$   & $0.000$ & $2.26$/$2.26$ \\
\hline
V$_\mathrm{Se}$@1L 	&$1.18$  &$-4.588$/$4.588$ &$4.583$/$-4.583$ &$-4.588$/$4.587$ &$-0.001$/$0.003$/$-0.001$ & $0.001$ & $1.79$/$1.79$\\
V$_\mathrm{Se2}$@1L 	&$1.50$  &$-4.586$/$4.586$ &$4.596$/$-4.595$ &$-4.593$/$4.592$ &$-0.001$/$0.003$/$0.000$  & $0.002$ &$1.26$/$1.26$\\  
V$_\mathrm{Se2}$@2L 	&$1.26$  &$-4.593$/$4.593$ &$4.579$/$-4.579$ &$-4.593$/$4.593$ &$0.000$/$0.000$/$0.000$   &$0.000$ & $1.75$/$1.75$ \\
\hline
\end{tabular*}
\end{table*}
\begin{figure*}[h]
\centering
\includegraphics[width=0.92\textwidth]{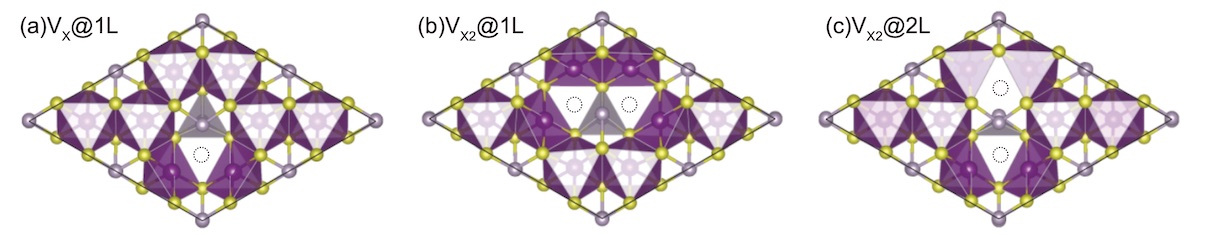}
\caption{MnPX$_3$ monolayer in the polyhedral representation to illustrate different vacancies configurations: (a) V$_\mathrm{X}$@1L, (b) V$_\mathrm{X2}$@1L, and (c) V$_\mathrm{X2}@$2L. Dotted circles indicate the respective chalcogen vacancies.}
\label{VacancyStructures}
\end{figure*}

It is well known that the optical properties strongly depend on the band structures. As mentioned above, band structures obtained by PBE+$U$ method show main features as that of HSE06 functional. Herein, we paid our attentions on the optical properties of 2D MnPX$_3$ evaluated by PBE+$U$ method. The lower panels of Fig.~\ref{2DBandStructure} (a) and (b) show the curves for $P^2$ with transitions from VB(1,2,3) to CB(1,2) (VB1 means the top VB, and VB2 the second top VB, etc.; see upper panels of Fig.~\ref{2DBandStructure}) along the high symmetry $k$-path according to the band structures. The absorption coefficient spectra $\alpha(\omega)$ in a wide energy range are presented in Fig.~\ref{OpticsAbsorption} (a) for 2D MnPS$_3$ and (b) for 2D MnPSe$_3$, respectively. 

The curves of $\alpha(\omega)$ for the two MnPX$_3$ monolayers demonstrate some similar character in the whole energy range due to their similar band structures. All the spectra exhibit anisotropic character between in-plane $xx$ ($yy$) components and out-plane $zz$ components, corresponding to the 2D monolayer anisotropy. The threshold energy of the absorption spectra occurs at around $2.50$\,eV and $1.84$\,eV for 2D MnPS$_3$ and 2D MnPSe$_3$, respectively, according to their direct band gaps as fundamental absorption edge. However, the $P^2$ spectra shows a zero value at $\mathrm{K}$ point, it means a forbidden transition between the direct band edge. Additionally, the value of $P^2$ spectra between VB1 and CB1 is not very large, especially for 2D MnPS$_3$, therefore, the absorption coefficient starts to increase very slowly at the fundamental band gap threshold. Nevertheless, with higher energy phonon incidence, the $\alpha(\omega)$ spectra increases sharply when the associated matrix elements are large and the transitions are allowed. For instance, around the $\Gamma$ point, $P^2$ spectra show the largest values for transitions from VB3 to VB1, and the $\alpha(\omega)$ spectra increase rapidly around the corresponding energy about $3.8$\,eV for 2D MnPS$_3$ while shows a sharp peak at $2.6$\,eV for 2D MnPSe$_3$. According to the different band gaps, it is evident that the $P^2$ spectra of 2D MnPSe$_3$ show much larger values than that of MnPS$_3$ in the according transitions and $k$ points (Fig.~\ref{OpticsAbsorption}). For example, with $3.2$\,eV photon induced, the absorption coefficient $\alpha(\omega)$ shows a value of $2\times10^{5}\,\mathrm{cm}^{-1}$ for 2D MnPSe$_3$, which is four times larger than $0.5\times10^{5}\,\mathrm{cm}^{-1}$ for 2D MnPS$_3$.  As a consequence, the 2D MnPSe$_3$ monolayer is a stronger absorption efficiency material than 2D MnPS$_3$, especially in the fundamental optical range.

The $\alpha(\omega)$ spectra of these two compounds were further calculated taking into account SOC (see Fig.~\ref{OpticsAbsorption}) and the obtained curves demonstrate the similar features as the spectra calculated on the basis of initial PBE+$U$ approach. In the low energy range under $\sim5.0$\,eV, the absorption spectra calculated with PBE+$\,U$+SOC and PBE+$\,U$ almost overlap  each other, while show some difference above $\sim5.0$\,eV. Noticeably, the peaks marked by red and blue arrows at $2.6$\,eV and $3.4$\,eV reduce apparently with comparison to the PBE+$U$ spectra for MnPSe$_3$ SOC-$\alpha(\omega)$, it is owing to the large energy-splitting for the SOC-VBs at the $\Gamma$ point, leading to the descending of transition probabilities. Accordingly, the SOC bands of 2D MnPS$_3$ do not show apparent energy-splitting at the $\Gamma$ point (Fig.~\ref{2DBandStructureSOC} and ESI$^\dag$, Fig.\,S2), and the peaks marked by arrows at $3.4$\,eV and $4.2$\,eV do not exhibit much differences between SOC-$\alpha(\omega)$ and PBE+$U$-$\alpha(\omega)$ spectra. In addition, the marked peak at $2.6$\,eV for 2D MnPSe$_3$ can be attributed to the electron transitions between the two highest band of VBs and two lowest bands of CBs, and it is followed by a down-step absorption caused by the 0.25 eV sub-gap between $2.5$\,eV and $2.75$\,eV in the CBs. Meanwhile, the marked peak at $3.4$\,eV is followed by a up-step for $\alpha_{xx}(\omega)$ of MnPS$_3$, and it can be attributed to the low DOS for the unoccupied orbitals between $3.2$\,eV and $3.4$\,eV.

To further explore the underlining electron transition mechanism of 2D MnPX$_3$, the $k_x-k_y$ constant energy cuts of spin-textures were extracted from the calculated band structures as shown in Fig.~\ref{SpinTexture2D} (the chosen energies for the cuts are marked for every panel and the energy differences between the low and high values are almost equal to the energy of the typical peaks in the absorption spectra mentioned above). All spin-textures are spin degenerated, caused by the robust intrinsic AFM state for both studied materials. 
Most importantly, the spin-textures show various circles, indicating the existence of electron pockets in the energy space for 2D MnPX$_3$, which will benefit the electron transitions. For simplicity, we discussed the spin-textures of MnPSe$_3$ monolayer to explore the transition possess. For 2D MnPSe$_3$, the centre electron pocket at $\Gamma$ point belongs to VB1 at $-0.25$\,eV, the six pockets located along the $\Gamma-\mathrm{M}$ path belong to VB2 at $-0.25$\,eV and VB3 at $-0.95$\,eV, whereas the empty pockets at $+2.4$\,eV belong to both CB1 and CB2. Particularly, the six unoccupied pockets connected together as a whole big pocket for CB2. As shown in Fig.~\ref{2DBandStructure}, $P^2_\mathrm{VB2-CB2}$ spectrum exhibits a maximum peak at the $\Gamma$ point, the $P^2_\mathrm{VB2-CB1}$ and $P^2_\mathrm{VB2-CB2}$ present some extreme values along $\Gamma-\mathrm{M}$ path. Moreover, both $P^2_\mathrm{VB3-CB1}$ and $P^2_\mathrm{VB3-CB2}$ show much larger values along the $\Gamma-\mathrm{M}$ path. Combination of the abundant electron pockets and large value of $P^2$ is advantageous to the electron transitions between the occupied VBs and empty CBs, leading to the marked peaks for the absorption coefficient of 2D MnPSe$_3$ as discussed. According to the similar spin-textures, this mechanism can also be applied for the description of the electron transitions in 2D MnPS$_3$. The optical absorption and electronic structures indicate that 2D MnPX$_3$ monolayers would exhibit good performance for photocatalytic water splitting as shown in the previous study.~\cite{Zhang:2016kra}

\subsection{Vacancy defects in 2D MnPX$_3$}

The effect of sulphur and selenium vacancies on the magnetic and optical properties of 2D MnPX$_3$ is also investigated. In all considered vacancy systems, $2\times2\times1$ supercell was built to avoid the interaction between two defects or defect-pairs with the distance more than $10$\,\AA. Three different amounts and geometric configurations of defects are discussed (Fig.~\ref{VacancyStructures}): (i) one vacancy at X1 site, named as V$_\mathrm{S}$@1L or V$_\mathrm{Se}$@1L with a defect concentration about $4.2$\%; (ii) two neighbouring vacancies at  sites X2 and X3 of the same chalcogen sub-layer  named as V$_\mathrm{S2}$@1L or V$_\mathrm{Se2}$@1L with a concentration about $8.3$\%; (iii) two neighbouring vacancies at sites X1 and X4 of different chalcogen sub-layers named V$_\mathrm{S2}$@2L or V$_\mathrm{Se2}$@2L (see Fig.~\ref{CrystalStructure} (c)). The S/Se vacancy formation energy was calculated to evaluated the energetic stability~\cite{Ovcharenko:2016fu}: $E_\mathrm{f}=\frac{1}{n}(E_\mathrm{vac}+n\mu_\mathrm{X}-4E_\mathrm{MnPX_3})$, where $n$ is the defect number, $E_\mathrm{vac}$ and $E_\mathrm{MnPX_3}$ are the energies of relaxed vacancy system and pristine cell, respectively. The chemical potential energy is calculated from the most stable state of chalcogen crystal, as $\mu_\mathrm{S}=-4.13$\,eV with $P2/c$ space group and $\mu_\mathrm{Se}=-3.50$\,eV with $P2_1/c$ space group. The vacancy formation energies are presented in Table~\ref{vacancytable}. All $E_\mathrm{f}$ for S/Se vacancies are small, which implies easy defect formation in experiments. According to the different electronegativity of sulphur and selenium, it is found that the $E_\mathrm{f}(\mathrm{V}{_\mathrm{S}})>E_\mathrm{f}(\mathrm{V}{_\mathrm{Se}})$ in the corresponding systems, and $E_\mathrm{f}(\mathrm{V}{_\mathrm{X2}@\mathrm{1L}})>E_\mathrm{f}(\mathrm{V}{_\mathrm{X2}@\mathrm{2L}})$ with the same defect account, revealing that the Se vacancy is more energetically favourable to produce than the S one, whereas a pair of vacancy is energetically unfavourable to produce in the same layer as $\mathrm{V}{_\mathrm{X2}@\mathrm{1L}}$ than in the different layers as $\mathrm{V}{_\mathrm{X2}@\mathrm{2L}}$. 

\begin{figure}[h]
\centering
\includegraphics[width=0.46\textwidth]{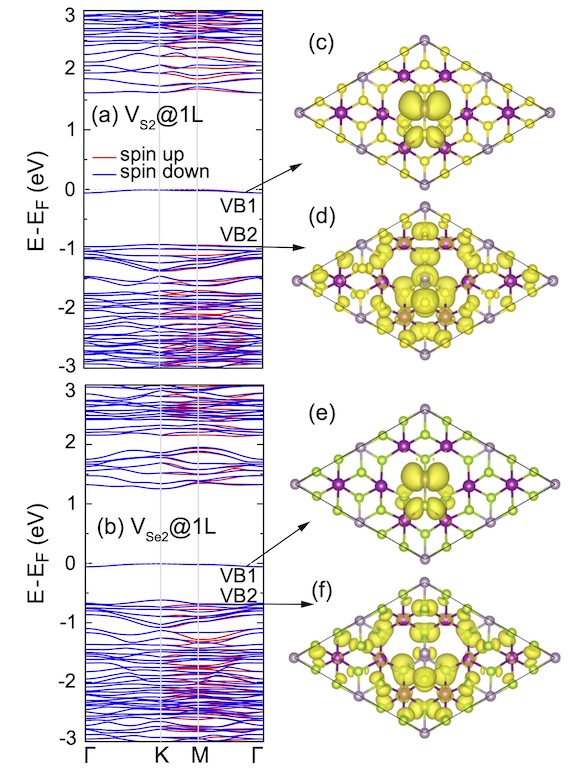}
\caption{Band structures of (a) 2D MnPS$_3$ V$_\mathrm{S2}$@1L and (b) 2D MnPSe$_3$ V$_\mathrm{S2}$@1L defect systems obtained with PBE+$\,U$. Partial charge densities (iso-surface value $=2\times10^{-3} e/$\AA$^3$) obtained for VB1 and VB2 of (c,d) 2D MnPS$_3$ V$_\mathrm{S2}$@1L and (e,f) 2D MnPSe$_3$ V$_\mathrm{S2}$@1L defect systems, respectively.}
\label{VacancyBand}
\end{figure}

The magnetic moments $M$ of different Mn ions, and total $M$ of $s,p$ or $d$ orbitals are listed in Table~\ref{vacancytable} for 2D vacancy MnPX$_3$ systems. Upon the vacancy defects induced, the 1X@1L and 2X@1L defect systems are all driven into ferrimagnetic state with a tiny net $M_\mathrm{total}$ about $0.001\,\mu_\mathrm{B}$ and $0.002\,\mu_\mathrm{B}$, respectively, whereas the 2X@2L vacancy systems remain AFM state without net magnetic moment. The different magnetic properties can be resulted from the changed system symmetry of vacancy systems. All the vacancy systems have a bilateral-symmetry axis along $\vec{a}+\vec{b}$ direction, while the 2X@2L system has one more inversion symmetry in the same direction. Accordingly, the Mn honeycomb lattice in a defect supercell shows some deformations with different $M$ value for each Mn ions, as listed in Table~\ref{vacancytable}. The total magnetic moments derived from different orbitals are also presented, and it should be noted that the net magnetic moment is dominated by the S/Se $p$ orbitals around defects. However, with such slight net $M_\mathrm{total}$, the influence of chalcogen defects on the 2D MnPX$_3$ monolayer can be almost neglected. In other words, the negligible $M_\mathrm{total}$ of vacancy MnPX$_3$ further demonstrate the strong AFM interactions between the Mn ions in the honeycomb lattice.

In order to seek the influence of chalcogen vacancy on the electronic properties of 2D MnPX$_3$, band structure, total and partial DOS of different defect-derived systems were then calculated (see ESI$^\dag$, Fig.\,S3). According to the crystal symmetry change and net magnetic moment, the bands show small spin-split along the $\mathrm{K}-\mathrm{M}-\Gamma$ path, but remain spin degenerate along the $\Gamma-\mathrm{K}$ path for V$_X$@1L and V$_\mathrm{X2}$@1L systems, while band structures remain spin degenerate along the whole high-symmetry $\mathrm{k}$-paths for V$_{X2}$@2L systems. Strikingly, there is no isolated defect state generated within the band gap in the band structure for V$_\mathrm{X}$@1L and V$_\mathrm{X2}$@2L derived systems, but such a state appear in the band structure of the V$_\mathrm{X2}$@1L vacancy system (Fig.~\ref{VacancyBand}). These defect states appear as flat bands and they are mostly derived from the chalcogen and phosphorus $p$ orbitals, as shown in the partial DOS. Herein, the partial charge density plots of VB1 and VB2 are illustrated in Fig.~\ref{VacancyBand} (c) and (d) for V$_\mathrm{S2}$@1L, (e) and (f) for V$_\mathrm{Se2}$@1L, respectively. As the chalcogen vacancy pair locates at X2 and X3 sites, the charge distribution mostly concentrates at neighbouring P1 and X1 ions, and partly at X5 and X6 sites for the mid-gap bands, all displaying $p$-like features, confirming the $p$ orbital contribution to the defect state. For the VB2 states, the electron distributions locate at vacancy-neighbouring ions more than on the other ions, revealing that there are some defect states locating just under the Fermi level. Due to the defect states, the band gaps of vacancy systems are smaller than that of pristine monolayers, as listed in Table~\ref{vacancytable}. 

Without mid-gap bands generated, the uppermost VBs apparently show some flat band behaviours for V$_\mathrm{X}$@1L and V$_\mathrm{X2}$@2L monolayers, as shown for their band structures (ESI$^\dag$, Fig.\,S3). It also can be further confirmed by the partial electron distribution which localises around the vacancies (ESI$^\dag$, Fig.\,S4). As a consequence, the defect states are mainly located just under the Fermi level for the vacancy MnPX$_3$ system without mid-gap state. From comparison, the defect states of 2D MnPS$_3$ vacancy monolayer are more localised than that of the corresponding 2D MnPSe$_3$ monolayer. Due to the defect states, the band gaps of vacancy systems are smaller than that of pristine monolayers, as listed in Table~\ref{vacancytable}. 

\begin{figure}
\centering
\includegraphics[width=0.45\textwidth]{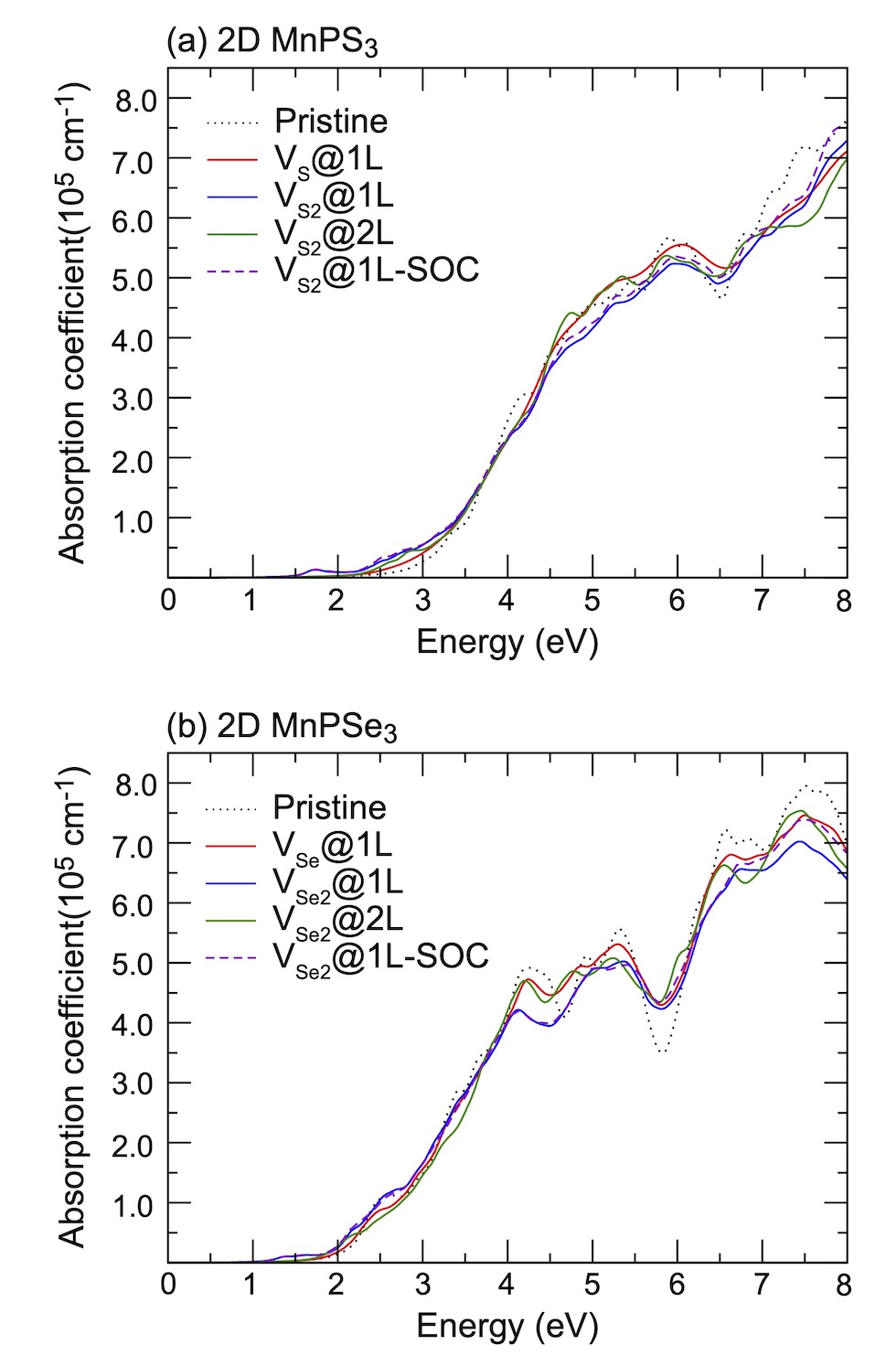}
\caption{The $xx$ component of the absorption coefficient $\alpha(\omega)$ as a function of energy for pristine and defected (a) 2D MnPS$_3$ and (b) 2D MnPSe$_3$ systems.}
\label{Optics_vacancies}
\end{figure}

\begin{figure}
\centering
\includegraphics[width=0.48\textwidth]{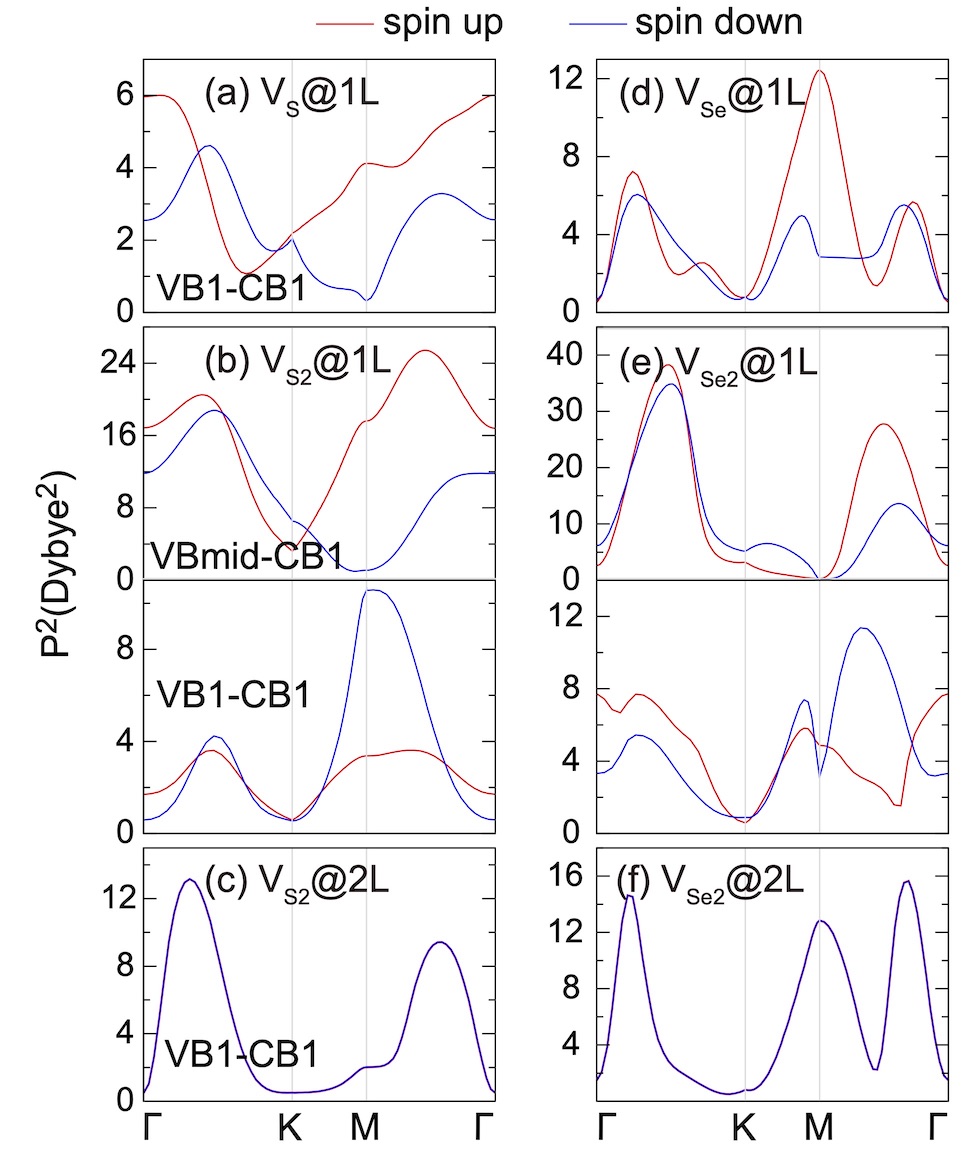}
\caption{The square of transition dipole moment matrix for 2D MnPS$_3$ (left panels) and 2D MnPSe$_3$ (right panels) defected systems.}
\label{Vacancy-TDM}
\end{figure}

Vacancy defects strongly influence the optical properties of the host material. Herein, the absorption spectra of defected MnPX$_3$ monolayers were investigated and the $xx$ components of $\alpha(\omega)$ are plotted in Fig.~\ref{Optics_vacancies}. The $\alpha(\omega)_{xx}$ spectra of vacancy systems show the similar characters as those of pristine monolayers in the whole energy range except for the new defect-related peaks. The extreme values of absorption spectra alternate between the different systems with the continuous incident photon energy, revealing almost equal absorption efficiency for both pristine and defected systems. Some new features for the absorption spectra of vacancy systems should be noticed. Caused by the reduced band gaps, the $\alpha(\omega)$ spectra threshold shifts to lower energy for the vacancy systems, compared with the $\alpha(\omega)_{xx}$ spectra of pristine monolayers. Particularly, a significant defect-peak appears at $1.70$\,eV for V$_\mathrm{S2}$@1L and $1.40$\,eV for V$_\mathrm{Se2}$@1L system, attributed to the direct transitions between the VB1 states to
the lower part CBs of these $n$-type semiconductors. Similarly, the defect induced peaks in the energy range from $2.5$\,($2.0$)\,eV to $3.0$\,($2.5$)\,eV for MnPS$_3$ (MnPSe$_3$) vacancy systems also show significant width, although the defect states form flat bands under the Fermi level for V$_\mathrm{X}$@1L and V$_\mathrm{X2}$@2L systems. The SOC effect on the absorption spectra were also considered into the vacancy systems. For simplicity, only the $xx$ component $\alpha(\omega)$ spectra of V$_\mathrm{X2}$@1L systems are presented. As that of pristine monolayers, the SOC-$\alpha(\omega)$ spectra also show more smooth than that of PBE+$\,U$-$\alpha(\omega)$ spectra. Absorption spectra usually can be used as criteria to assess the crystal quality of the monolayers. Here, only the V$_\mathrm{X2}$@1L vacancy can be easily assessed with a new peak within the band gap for the absorption spectra. In addition, both
V$_\mathrm{X}$@1L and V$_\mathrm{X2}$@1L vacancies could be probed by circularly polarised light due to the spin-splittings, as determined by the different transition dipole moments which show large splits between the spin up and spin down channels (Fig.~\ref{Vacancy-TDM}). However, the V$_\mathrm{X2}$@2L defects cannot be easily detected because of the absence of both, in-gap absorption peak and spin-splittings. From the discussion above, some further theoretical and experimental studies on the chalcogen vacancy defects should be carried out, such as the circularly polarisation, bound exciton and net magnetic detection, to identify the defect states and influence to the optical and magnetic properties of 2D MnPX3 monolayers.

\section{Conclusions}

A systematical first-principle studies on the electronic, magnetic and optical properties of 2D transition-metal phosphorous trichalcogenide MnPX$_3$ (X = S, Se) were carried out based on the density functional theory. The bulk MnPS$_3$ in the $C2/m$ space group and MnPSe$_3$ in the $R\overline{3}$ one showing AFM semiconductor behaviour with direct band gap. For both materials, the monolayer form is energetically favourable and these layers can be exfoliated from the bulk phase with a small cleavage energies, as $0.12$\,J/m$^2$ for MnPS$_3$ and $0.23$\,J/m$^2$ for MnPSe$_3$, respectively, which are much lower than $0.37$\,J/m$^2$ of graphite. Confirmed by the phonon spectrum with no imaginary dispersion, MnPX$_3$ monolayers show a good dynamical stability. The 2D MnPX$_3$ monolayers are N\'eel AFM semiconductors with a direct band gap value of $2.37$\,eV (PBE+$U$) or $3.08$\,eV (HSE06) for 2D MnPS$_3$ and $1.84$\,eV (PBE+$U$) or $2.56$\,eV (HSE06) for 2D MnPSe$_3$ at the K point, in excellent agreement with the experimental data. The NN, 2NN and 3NN exchange parameters are all positive, revealing the AFM-N\'eel ground state of 2D MnPX$_3$. Using periodic boundary conditions, Monte Carlo simulations gave the theoretical $T_\mathrm{N}$ of $103$\,K and $80$\,K for 2D MnPS$_3$ and MnPSe$_3$, respectively. With high spin state of Mn ions arranged in a honeycomb lattice, 2D MnPX$_3$ shows valley polarisation with spin-degeneracy in the band structure if the spin-orbital coupling is considered. Moreover, in-plane strain offers an undesirable effect for the valley polarisation. With direct band gaps falling into the visible optical spectrum, MnPX$_3$ monolayers have good performance on the optical absorption which were investigated based on the electronic structures and transition dipole moment matrix. The influence of a single and a pair of chalcogen vacancies on the electronic, magnetic and optical properties of MnPX$_3$ were also investigated, and the effect is strongly correlated with the vacancy structure configurations. Two vacancies in the same chalcogen sublayer will introduce mid-gap state and spin-splitting, whereas two vacancies in different chalcogen sublayers show no mid-gap states and no spin-splitting. From the absorption spectra of vacancy systems, it is proposed that optical absorption spectra cannot be used as an ideal criteria to determine the crystal quality of the 2D MnPX$_3$ monolayers. 

\section*{Conflicts of interest}
There are no conflicts to declare.

\section*{Acknowledgements}
This work was supported by the National Natural Science Foundation of China (Grant No. 21973059). Y.\,J. thanks the support from the Natural Science Foundation of Hubei Province (Grant No. 2018CFB724) and of Education Department (Grant No. D20171803). E.\,V. thanks the support by the Ministry of Education and Science of Russian Federation within the framework of the State Assignment for Research, Grants no. 4.6759.2017/8.9. 

\bibliography{references_all.bib}

\clearpage

\noindent Supplementary Information for: Electronic, magnetic and optical properties of MnPX$_3$ (X = S, Se) monolayers with and without chalcogen defects: A first-principle study

\clearpage

\begin{figure*}
\begin{center}
\includegraphics[width=0.75\textwidth]{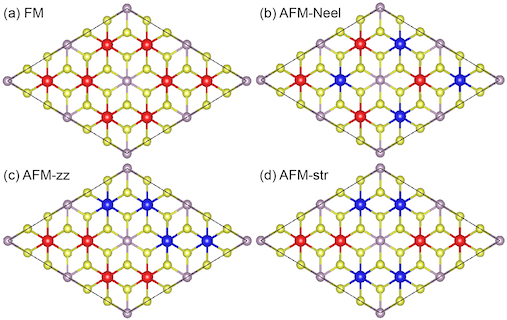}
\end{center}
\textbf{Fig.\,S1.} Four different magnetic configurations of MnPX$_3$: (a) FM, (b) AFM-N\'eel, (c) AFM-zig-zag, and (d) AFM-stripy. Up and down spin moments orientations for Mn ions are coded by red and blue spheres, respectively.
\label{MagneticStructures}
\end{figure*}

\begin{figure*}
\begin{center}
\includegraphics[width=0.98\textwidth]{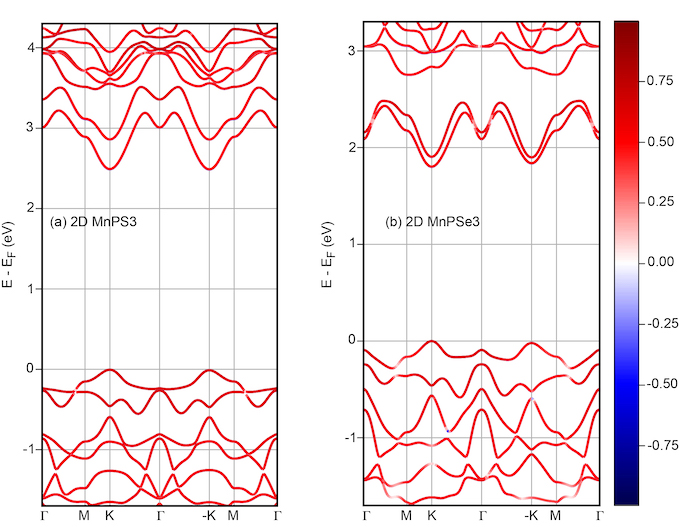}
\end{center}
\label{SOCbands-SI}
\textbf{Fig.\,S2.} Band structures of (a) 2D MnPS$_3$ and (b) 2D MnPSe$_3$ with SOC taken into account. The red and blue colors denote the spin up and down states, respectively.
\end{figure*}
\newpage

\clearpage
 
\begin{figure*}
\begin{center}
\includegraphics[width=0.9\textwidth]{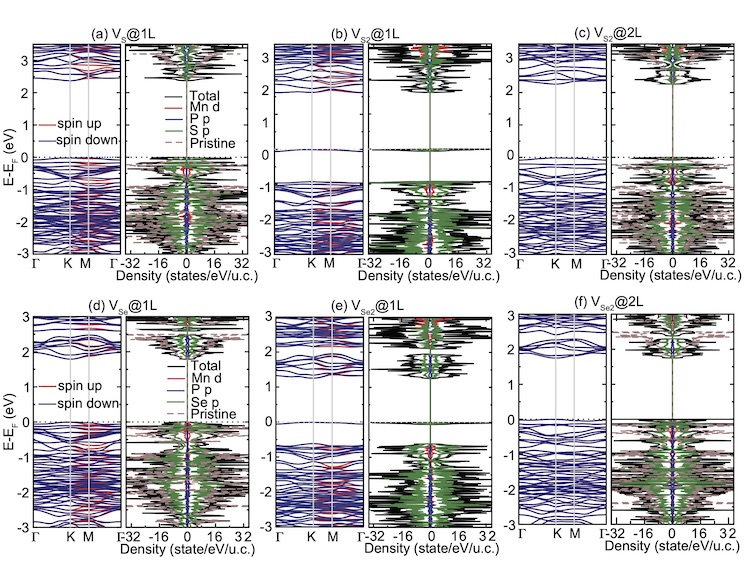}
\end{center}
\textbf{Fig.\,S3.} Band structures, total and partial DOS for 2D MnPS$_3$ (upper panels) and 2D MnPSe$_3$ (lower panels) vacancy systems.
\label{VacancyBand-SI}
\end{figure*}

\clearpage

\begin{figure*}
\begin{center}
\includegraphics[width=0.75\textwidth]{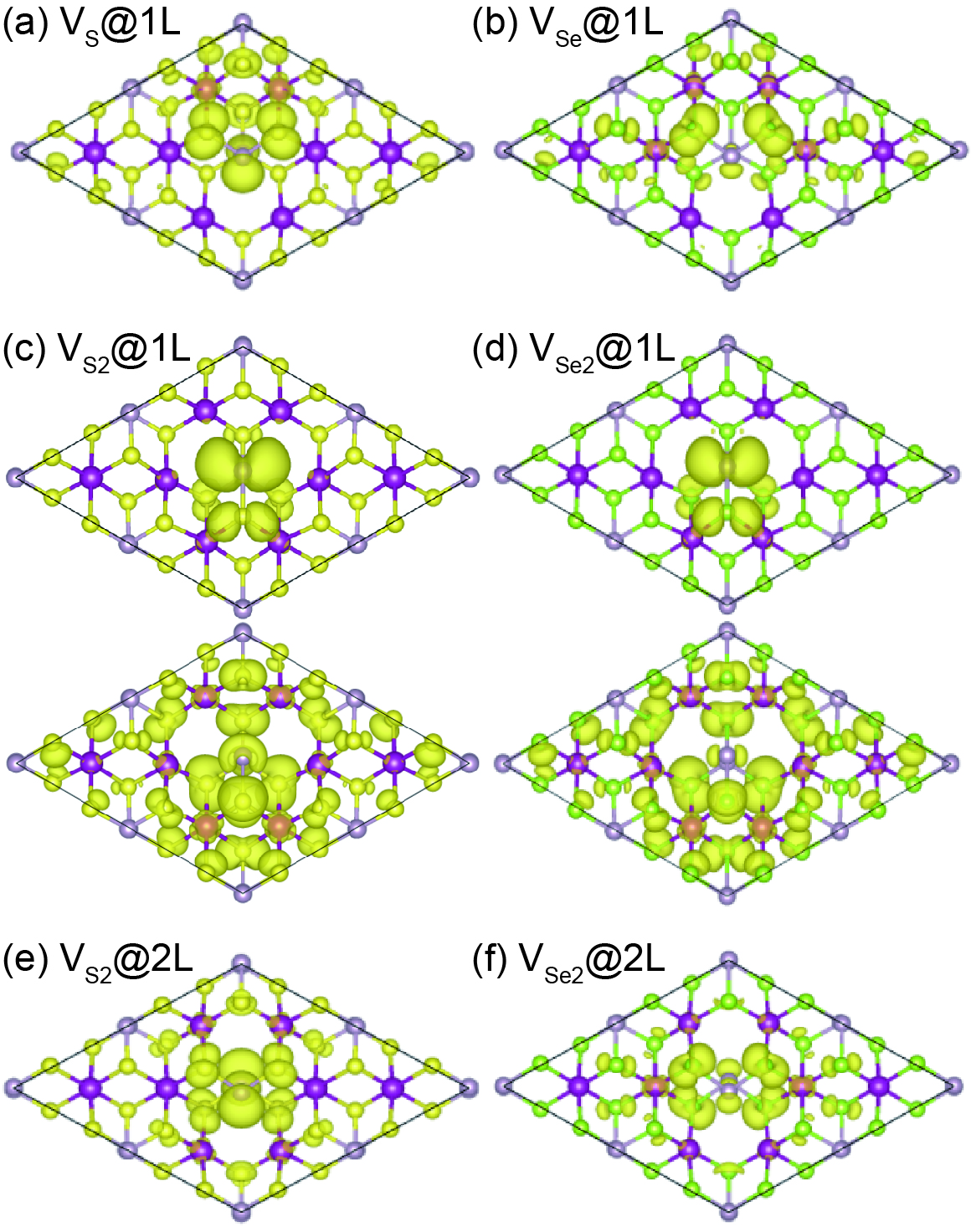}
\end{center}
\textbf{Fig.\,S4.} Partial charge densities (iso-surface value $=2\times10^{-3} e/$\AA$^3$) of 2D MnPS$_3$ (left panels) and 2D MnPSe$_3$ (right panels) vacancy systems.
\label{ChargeDensity-SI}
\end{figure*}

\end{document}